\documentclass[a4paper,12pt]{article}
\usepackage{amstex,righttag,epsfig,rotating}
\newcounter{num}[section]
\newenvironment{theorem}%
{\addtocounter{num}{1}\begin{flushleft}\textbf{Theorem \thesection
.\thenum:\ }} 
{\end{flushleft}}
\renewcommand{\theenumi}{\roman{enumi}}

\newcommand{\scs}{\scriptsize}
\newcommand{\pic}[3]{
\begin{minipage}{6.4cm}
\setlength{\unitlength}{1.28cm}
\begin{picture}(5,3.3)
\epsfxsize=6cm
\put(0.2,0.1) {\epsfbox{#1}}
\put(0.1,1.7){#2}
\put(2.8,0.1){#3}
\end{picture}
\end{minipage}
}
\newcommand{\spic}[2]{\pic{#1}{$\scriptstyle #2$}{$\scriptstyle t$}}
\newcommand{\picgen}[7]{
\begin{minipage}{6.2cm}
\setlength{\unitlength}{1.28cm}
\begin{picture}(4.8437,3.3)
\epsfxsize=6cm
\put(0.2,0.1) {\epsfbox{#1}}
\put #2
\put #3
\put #4
\put #5
\put(0.1,1.7){#7}
\put(2.8,0.1){#6}
\end{picture}
\end{minipage}
}
\newcommand{\picth}[9]{
\picgen{#1}{#2 {$\scriptstyle #3$}}
{#4 {$\scriptstyle #5$}}{#6 {$\scriptstyle #7$}}{#8 {$\scriptstyle #9$}}
{$\scriptstyle t$}{$\scriptstyle \theta \cdot t$}
}

\begin{document}
\begin{titlepage}
\begin{flushright}
gr-qc/9802043
\end{flushright}
\begin{center}
\vfill
{\large\bf Evolution of the Bianchi I, the Bianchi III and the Kantowski-Sachs
Universe: Isotropization and Inflation$^*$}
\vfill
{\bf Samuel Byland$^{1}$ and David Scialom$^{2}$}
\vskip 0.5cm
$^1$ Institute of Theoretical Physics, University of Z\"{u}rich, 
\\Winterthurerstrasse
190, CH-8057 Z\"{u}rich, Switzerland\\
$^2$ D\'epartement d'Astrophysique Relativiste et de Cosmologie,
CNRS-Observatoire de Paris, 92195 Meudon, France
\end{center}
\vfill
\begin{abstract}
\noindent We study the Einstein-Klein-Gordon equations for a convex
positive potential in a Bianchi I, a Bianchi III and a Kantowski-Sachs
universe. After analysing the inherent properties of the system of
differential equations, the study of the asymptotic behaviors of the
solutions and their stability is done for an exponential potential.  
The results are
compared with those of Burd and Barrow. 
In contrast with their results, we show that for the
BI case isotropy can be reached without inflation and we find new
critical points which lead to new exact solutions. On the other hand 
we recover the
result of Burd and Barrow that if inflation
occurs then isotropy is always reached.
The numerical integration is also done and all the asymptotical
behaviors are confirmed.  
\end{abstract}

\noindent PACS number(s): 98.80.Cq, 98.80.Hw
\vfill
$^*$ This work was partially supported by the Swiss National Science
Foundation.
\end{titlepage}

\section{Introduction}\label{secINT}
Inflation, first introduced by Guth \cite{Gut:81}, was introduced
in the standard cosmological model to solve the homogeneity, the isotropy
and the horizon problem. The latter is well explained due to the fact
that inflation is characterized by an exponential or power-law
expansion of the universe and at the same time a quasi-constant
behavior of the Hubble horizon. 

On the other hand the homogeneity and the isotropy
problem is in fact not well explained because from the start the
homogeneous and isotropic Friedman-Lema\^{\i}tre metric is used. 
To really solve the problem one should start with an arbitrary metric
and show that inflation takes place and that the universe evolves towards
a Friedman-Lema\^{\i}tre metric. The problem of the onset of inflation
was considered numerically for spherical inhomogeneous cosmologies
\cite{Gol:89,Gol:90} and a semi-numerical analysis was done for inhomogeneous,
quasi-isotropic universes \cite{Der:95} using the long wavelength
iteration scheme \cite{Bel:72,Com:94}. They showed that a large
initial inhomogeneity supresses the inflation stage.
Because of the analytical difficulties of the
task, one can as a first step, consider only a homogeneous but
anisotropic metric and try to solve the isotropy problem. 

This task was first undertaken by Collins and Hawking \cite{Col:73} who
showed that, within the Bianchi type universe filled by matter
satisfying the dominant energy condition and positive pressure
criterion, the isotropy problem can only be solved for the types I, V,
VII$\hbox{}_{\hbox{o}}$ and VII$\hbox{}_{\hbox{h}}$. 
They showed also that only a subclass of vanishing measure
in the space of all homogeneous initial conditions can approach
isotropy. 

With the presence of an inflationary stage, when the dominant energy
condition is violated, there was hope to obtain a cosmic no-hair theorem. 
The study  done by Heusler \cite{Heu:90} showed there is no
no-hair theorem for a real scalar
field having a convex positive potential with a vanishing local minimum in a
Bianchi type universe. 
In fact isotropy can only be approached if the underlying Lie
group of the Bianchi type metric is compatible with a Friedman-Lema\^{\i}tre
model. 

Beside the Bianchi type metrics the Kantowski-Sachs model 
also describes a spatially homogeneous universe. This model with
a perfect fluid description of matter and with or without a cosmological
constant has
been studied by a number of authors
\cite{Col:77,Lor:82,Web:84,Gro:86,Gro:87}.  They found an anisotropic
asymptotical behavior of the model. Burd and Barrow \cite{Bur:88}
analyzed this system with a real scalar field having an exponential
potential as a source. Among others they found an anisotropic
asymptotical behavior or that  
inflation can  occur depending on the value of the coupling constant
entering in the definition of the potential. The exponential potential
is motivated by the fact that it can be obtained, for example, by dimensional
reduction of more fundamental theories \cite{Hal:87} or in conformal
equivalent theories of gravity whose Lagrangian is an arbitrary
analytic function of the scalar curvature \cite{Whi:84,Bar:88,Bic:74}.  

In this work, we will consider a real scalar field with a convex
positive potential, but not necessarily with a local minimum, in a 
Bianchi I, a Bianchi III and a Kantowski-Sachs model. 
These three model are in fact
very closely related since it will be shown in section \ref{secBAS} that
the differential 
equations describing the evolution of the models are the same and that
only the constraint equations differ. In section \ref{secINH}
we study the inherent properties of the differential equations and
we recover in the case of a local minimum of the potential a
result of Heusler \cite{Heu:90}. We also show why the solution of
the dynamical system must be studied for a potential having a
vanishing potential at infinity. Section \ref{secSTU} is devoted 
to the special case of an exponential potential. The result obtained
will be compared with the analysis done in Ref. \cite{Bur:88}. 
In section \ref{secNUM} the numerical integration of the system is done
and in section \ref{secCON} the results will be summarized.
\section{Basic equations}\label{secBAS}
We consider a scalar field with a convex positive potential $V$ in a
homogeneous 
universe having the following metric
\begin{equation}
	ds^2=g_{\mu\nu}\theta^{\mu}\theta^{\nu}=-\theta^0\theta^0+
	\delta_{ij}\theta^i\theta^j, \label{metr}
\end{equation}
where 
\begin{equation*}
	\theta^0=dt \, , \, 
	\theta^{1}=a(t)\,dr \, , \,
	\theta^{2}=b(t)\,d\vartheta \, , \,
	\theta^{3}=b(t)f(\vartheta)d\phi\, ,
\end{equation*}
with $f(\vartheta)=\vartheta$,  $f(\vartheta)=sinh(\vartheta)$ or
$f(\vartheta)=sin(\vartheta)$. We have respectively a Bianchi-I (B-I),
a Bianchi-III (B-III), and a Kantowski-Sachs (KS) metric. 

The action is given by
\begin{equation}
S=\int \left(-\frac{R}{16\pi G}+\frac{1}{2}e_\mu\varphi e^\mu\varphi
+ V(\varphi)\right)\sqrt{-g}\,d^4x \, ,
\end{equation}
where $g$ is the determinant of the metric and $e_\mu$ is the dual
basis of $\theta^\mu$. By varying the action with respect to the metric
we get the Einstein equations
\begin{gather}
2H_aH_b+H_b^2+\frac{k}{b^2}=8\pi G
\left(\frac{1}{2}\dot{\varphi}^2+V(\varphi)
\right)\, ,\label{e0}\\
2\dot{H_b}+3H_b^2+\frac{k}{b^2}=8\pi G
\left(-\frac{1}{2}\dot{\varphi}^2+V(\varphi) \right)\, ,\label{e1}\\
\dot{H_a}+\dot{H_b}+H_a^2+H_aH_b+H_b^2=8\pi G
\left(-\frac{1}{2}\dot{\varphi}^2+V(\varphi) \right)\, , \label{e2}
\end{gather}
with
\begin{equation}
H_a=\frac{\dot{a}}{a}\, ,\, H_b=\frac{\dot{b}}{b}\, ,\,
k=-\frac{1}{f}\frac{d^2f}{d\vartheta^2}\, .
\end{equation}
For the B-I, the B-III and the KS cases, we get respectively $k=0,-1, 1$.

The Klein-Gordon equation is obtained by varying the action with respect
to the scalar field. We get
\begin{equation}
\ddot{\varphi}+\left(H_a+2H_b\right)\dot{\varphi}+\frac{dV}{d\varphi}=0
\, .
\label{kg} 
\end{equation}

In the following we will express every quantity in units of the Planck
mass $m_p=1/\sqrt{8\pi G}$. This can be achieved by setting $8\pi G=1$
in the Einstein equations. The system is fully determined by the
independent equations (\ref{e0}), (\ref{e1}) and (\ref{kg}). One can
easily show that equation (\ref{e2}) follows from the others. This
is a reflection of the Bianchi identities. 
After some algebraic manipulations, these equations can be written as
a set of four first order differential equations, which are $k$
independent, and a constraint (eq.(\ref{e0})) which is conserved in
the evolution.

This set of equations can also be written in function of the 
expansion rate $\theta$ and the shear tensor $\sigma_{\mu\nu}$ of
the hypersurface of constant time $\Sigma$. We get for the first order
equations
\begin{align}
\dot{\theta} &= -\frac{1}{3}\theta^2-2\sigma^2+V(\varphi) -\psi^2\, ,
      \label{ed1}\\ 
\dot{\sigma} &= -\frac{1}{3\sqrt{3}}\theta^2-\theta\sigma +
      \frac{1}{\sqrt{3}} \sigma^2 +
      \frac{1}{\sqrt{3}}\left(V(\varphi)+\frac{1}{2}\psi^2\right)\, ,
      \label{ed2}\\  
\dot{\varphi} &=\psi\, ,\label{ed3}\\
\dot{\psi} &= -\theta\psi-\frac{dV}{d\varphi} \, ,\label{ed4}
\end{align} 
where $\sigma=\frac{1}{2}\:\sigma_{\mu\nu}\sigma^{\mu\nu}=\sqrt{1/3}(H_a-H_b)$ and 
$\theta=H_a+2H_b$. $\psi$ is defined by eq. (\ref{ed3}). As for the
constraint equation, it reads
\begin{equation}
\frac{1}{3}\theta^2+\frac{k}{b^2}=\sigma^2+V(\varphi)+\frac{1}{2}\psi^2\,
. \label{constraint}
\end{equation}
The solutions of eqs.(\ref{ed1})-(\ref{ed4}) do not
depend on the type of the homogeneous model considered. Since
eq.(\ref{constraint}) is conserved, we have to specify the homogeneous
model only in the initial conditions. In
the four dimensional space with coordinates $(\theta\:, \sigma\:,
z=\sqrt{V}\:, \psi )$ the B-I solutions are on the ``lightcone'', the
B-III ones are in it and the KS solutions remains outside the
lightcone (see Fig. 1). 

The advantage of taking these geometrical meaningful
quantities is obvious:
the shear tensor expresses the direction dependent
deviation from the   
global expansion. Hence, $\sigma$ measures the anisotropy. As a
criterium of isotropization we will not use the vanishing of $\sigma$
(as in \cite{Men:91}), but a stronger condition \cite{Col:73,Bur:88}
which is 
\begin{equation}
\frac{\sigma}{\theta}\rightarrow 0\: , \mbox{\hskip 1cm as\hskip 1cm}
t\rightarrow\infty\: . 
\end{equation}
In this paper, we want to analyze the evolution of an expanding
universe. Hence, we will restrict ourselves to positive values of
$\theta$. We want to know under which conditions the isotropization of 
this model is generic and when inflation occurs. 
In some cases, general properties of the solutions can be found  by
just analyzing the differential equations.

\section{Inherent properties of the dynamical system}\label{secINH}
This section generalizes, in some respect, a result obtained by
M. Heusler \cite{Heu:90}. 
In the B-I ($k=0$) and the B-III ($k=1$) case,  the asymptotic
behavior of the solutions can be obtained directly by the study of
eqs.(\ref{ed1})-(\ref{constraint}). Since we are interested in the
evolution of an expanding universe, we suppose that $\theta$ is
positive or equal to zero at
some time $t_0$. We also restrict ourselves to positive and convex
function $V$. From eq.(\ref{ed1}) and eq.(\ref{constraint}), 
we get the following inequality relation ($k\leq 0$)
\begin{equation}
\dot{\theta}\leq V-\frac{1}{3}\theta^2\leq 0\: . \label{in1}
\end{equation} 
Using the last inequality relation and again
eqs.(\ref{ed1},\ref{constraint}), it is easy to obtain
\begin{equation}
\dot{\theta}+\theta^2 =3V-\frac{k}{b^2}\geq 0\: . \label{in2}
\end{equation}
Setting $\mathcal{V}=ab^2$ which is positive, we can write
\begin{equation}
\theta=\frac{\dot{\mathcal{V}}}{\mathcal{V}}\: . \label{deft}
\end{equation}
By hypothesis, $\dot{\mathcal{V}}$ is positive at $t_0$.
Substituting eq.(\ref{deft}) in eq.(\ref{in2}), we have that
$\ddot{\mathcal{V}}/\mathcal{V}\geq 0$. Therefore, $\dot{\mathcal{V}}$ can only increase 
with time and
$\theta$ will remain positive after $t_0$. We have
shown that $\theta$ is a positive, monotonic decreasing
function and thus, we must have 
\begin{equation}
\dot{\theta}\rightarrow 0\mbox{\hskip 1cm and\hskip 1cm}
\theta\rightarrow\theta_\infty\geq 0\mbox{\hskip 1cm as\hskip 1cm}
t\rightarrow\infty\: . 
\end{equation}
This function converges towards the critical value $\theta_\infty$. It
is also easy to get
\begin{equation}
\dot{\theta}=\frac{k}{b^2}-3\sigma^2-\frac{3}{2}\psi^2\leq 0\: .
\end{equation}
Since the l.h.s of the inequality is a sum of negative terms, we have that
each of them must vanish as $t\rightarrow\infty$. Using this fact in
the constraint equation, we obtain
\begin{equation}
\theta\rightarrow\sqrt{3V}\mbox{\hskip 1cm as\hskip 1cm}
t\rightarrow\infty\: . 
\end{equation}
Eq.(\ref{ed4}) describes a damped harmonic oscillator, so we know
that the system must ``land'' at the minimum of the potential $V_0$
and thus 
\begin{equation}
\theta_\infty=\sqrt{3V_0}\: .
\end{equation}
For an exponential potential, the only minimum is at infinity and is
value is zero. Let us summarize the results by the following theorem
\begin{theorem}
Let be $k=0$ or $k=-1$ (Bianchi-I and Bianchi-III case). If at a given
time $t_0$ we have an expanding universe, that is $\theta(t_0)\geq 0$, then
\begin{enumerate}
\item $\dot{\theta}(t)\leq 0$, $\theta(t)\geq 0$ for all $t\geq t_0$, 
\item $\sigma(t)$, $\psi(t)$, $\frac{k}{b(t)^2} \rightarrow 0$,
for $t\rightarrow\infty$ and
\item $\theta\rightarrow\sqrt{3V_0}$, where $V_0$ is the minimum
of the potential and in particular $\theta\rightarrow 0$ for an
exponentially potential (i.e. $V = V_0 e^{-\lambda\varphi}$) for
$t\rightarrow\infty$.
\end{enumerate}
\end{theorem}
This theorem restates the proposition 1 in \cite{Heu:90} except than it
can now also be applied for a convex potential $V$ having its minimum at
infinity. 
If $V_0 > 0$, we can use the known result that the B-I and the B-III
converge exponentially to the isotropic De-Sitter universe \cite{Wal:83}. 
If $V_0 = 0$ at some finite value $\varphi_0$ then we only have 
isotropization in the B-I case \cite{Heu:90}. 

The above theorem gives no answer on isotropization for
an exponential potential in a B-I or a B-III model. Indeed, $\theta$
and $\sigma$ vanish together. In this case, the asymptotic
solutions of the dynamical system (eqs(\ref{ed1})-(\ref{ed4})) must be
found to know under which conditions isotropy and/or inflation can
occur. Since the equations of the dynamical system do not depend on
$k$, the KS case will also be solved. This system has been actually
studied in a paper of Burd and Barrow \cite{Bur:88}. 
However, their analysis 
on the system was not complete. Their analysis covers a part of the phase
space only. Indeed, only  two of the critical
points were found and some conclusions turn out to be incorrect. 

In the next section, the analysis of the exponential potential case
will be done in detail and we will compare our results with
Ref. \cite{Bur:88}. 

\section{Study of the dynamical system with an exponential
potential}\label{secSTU} 

\subsection{The singular points}

For a qualitative discussion of the system of differential equations
\eqref{ed1}-\eqref{ed4} we have to determine the asymptotic behavior near the
critical points. The usual methods of treating the problem rapidly turn
out to be inadequate for this case: the singular points are highly
non-hyperbolic so that the linearization or even the transformation to a
normal-form lead nowhere.   
Usually, the constraint equation can be used as a
Ljapunov-function. But in this case, eq. \eqref{constraint} has no
isolated root and therefore is of no help. There is no other obvious
candidates for a Ljapunov-function. 

The problems originate from the fact that there are only quadratic terms
on the right hand side of eqs.\eqref{ed1}-\eqref{ed4}. There is a standard
procedure 
for such systems: one has to ``divide'' all variables but one by the
remaining variable and rewrite the system in terms of these new variables. Of
course this transformation is only non-singular if the divisor does not
vanish in the whole region of definition. In Ref. \cite{Bur:88}, the
remaining variable was chosen to be $y=V(\Phi)$, 
which does not vanish in finite regions
of the phase space. Here, we prefer to distinguish the variable
$\theta$ since it will allow us to find additional critical points (in
finite regions of phase space). This transformation is
well defined outside the origin.

Defining the variables $S$, $U$ and $P$ by
\begin{equation}
\sigma=S\, \theta, \ z=U\, \theta, \ \psi=P\, \theta
\end{equation}
and the time $\tau$ by
\begin{equation}
'=\frac{d}{d\tau}=\frac{1}{\theta}\frac{d}{dt}
\end{equation}
the differential equations \eqref{ed1}-\eqref{ed4} transform into
\begin{gather}
\theta' = \theta (-\frac{1}{3}-2 S^2+U^2-P^2) \label{transSysa}\\
S' =-\frac{1}{3 \sqrt{3}}-\frac{2}{3} S+\frac{1}{\sqrt{3}} S^2 +
\frac{1}{\sqrt{3}} U^2+\frac{1}{2 \sqrt{3}} P^2+2 S^3-S U^2+S
P^2\label{3eq1}\\
 U'= U (\frac{1}{3}-\frac{\lambda}{2} P+2
S^2-U^2+P^2)\label{3eq2}\\  
P' =-\frac{2}{3} P+\lambda U^2+2 S^2 P-U^2 P+P^3 \label{3eq3} \\
\intertext{and the constraint equation is}
S^2+U^2+\frac{1}{2} P^2+\frac{k}{b^2} = \frac{1}{3}\ . \label{transSysConst}
\end{gather}
The dynamical system defined by eqs.\eqref{3eq1}-\eqref{3eq3},
which are independent of $\theta$, will be referenced from now on by
the tag ($*$). 

For the Bianchi I case ($k=0$), equation \eqref{transSysConst} describes
the surface of an ellipsoid, say $E$, which separates the
Kantowski-Sachs (inside) 
from the Bianchi~III (outside) solutions.
The Bianchi~I model ($k=0$) has to be treated carefully because in this
case the system can only evolve on a twodimensional submanifold $W$ in
the space of the variables $(S,U,P)$, defined by
\begin{equation}
W=f^{-1}(\{\frac{1}{3}\}) \quad \text{with
$f(S,U,P)=S^2+U^2+\frac{1}{2}P^2$}\ .
\end{equation}
It may happen that $W$ coincides with the stable manifold of a critical
point which is unstable with respect to the whole system ($*$). In
this case the point is nevertheless stable for the Bianchi~I metric.

We will concentrate on determining the asymptotic behavior of this reduced
system. $\theta$ is then given by a simple integration of
\eqref{transSysa}.
After a short calculation we find for the system ($*$) three
discrete critical points and a one-parameter family of critical points:
\begin{align}
P_1: \quad & S = -\frac{1}{2 \sqrt{3}}, \ U = 0, \ P = 0
\label{kritP1}\\ 
P_2: \quad & S = 0, \ U =
\sqrt{\frac{6-\lambda^2}{18}}, \ P = \frac{\lambda}{3} \quad
\text{for $\lambda \leq \sqrt{6}$} \label{kritP2}\\ 
P_3: \quad & S =
\frac{1}{2 \sqrt{3}} \frac{2-\lambda^2}{\lambda^2+1}, \ U =
\frac{\sqrt{\lambda^2+2}}{\sqrt{2} (\lambda^2+1)}, \ P =
\frac{\lambda}{\lambda^2+1} \label{kritP3} \\ 
\Sigma: \quad & U = 0, \ 3
S^2+\frac{3}{2} P^2 = 1 \quad \text{with $S \in
[-\sqrt{1/3},\sqrt{1/3}]$} \label{kritS}\ .
\end{align}
Comparing these critical points with the results of Burd and Barrow
\cite{Bur:88} one easily verifies that $P_2$ and $P_3$ correspond to
their critical points $(IV)$ and $(III)$ whereas $P_1$ and the points of
$\Sigma$ have no counterpart in their paper. 
Using the variables in Ref. \cite{Bur:88}, $P_1$ and $\Sigma$ can be
found as critical points lying at infinity in their phase space. 

To be sure that there are no additional critical points at infinity in
our coordinate system  either, we first introduce spherical coordinates
\begin{equation}
S=r\, \sin\vartheta,\ U=r\,\cos\varphi\,\cos\vartheta,\
P=r\,\sin\varphi\,\cos\vartheta
\end{equation}
and then map the points at infinity ($r=\infty$) to the surface of a
unit sphere by the transformations
\begin{equation}
r \longrightarrow \rho=\frac{r}{r+1},\ d\tau \longrightarrow
d\eta=\frac{d\tau}{1-\rho}\ .
\end{equation}
At infinity ($\rho=1$) we thus find
\begin{align}
\frac{d\rho}{d\eta} &= 2-\cos^2\!\vartheta(1+2
\cos^2\!\varphi)\label{infeq1}\\ \frac{d\varphi}{d\eta} &=
\frac{\lambda}{2} \cos\varphi \cos{\vartheta} (2
\cos^2\!\varphi+\sin^2\!\varphi) \label{infeq2}\\
\frac{d\vartheta}{d\eta} &= \frac{\cos\vartheta}{2 \sqrt{3}}(2 -
\cos^2\!\vartheta + \cos^2\!\varphi \cos^2\!\vartheta-\sqrt{3} \lambda
\sin\varphi \cos^2\!\varphi \sin\vartheta \cos\vartheta)\
.\label{infeq3}
\end{align}
For a critical point at infinity, expressions \eqref{infeq1} --
\eqref{infeq3} have to vanish simultaneously which is easily seen to be
impossible. Therefore the points $P_1$, $P_2$, $P_3$ and the points of
$\Sigma$ are really the only critical points of the system ($*$).

It is worthwhile to note that the critical points correspond
to exact solutions of our original system
(\eqref{ed1}-\eqref{ed4}). Indeed, if we consider the
pull-back of a critical point ($S_k,U_k,P_k$) 
of ($*$) for a non-constant potential ($\lambda>0$), the volume
expansion $\theta$ is  given by
\begin{equation}
\theta=\theta_0 e^{A_k \tau}
\end{equation}
with $A_k=-1/3-2 S^2+U^2-P^2|_{(S,U,P)=(S_k,U_k,P_k)}$, and as a function of
the original time $t$ we obtain the following exact solutions
\begin{equation}\label{BackTr}
\theta=-\frac{1}{A_k\,t},\ \sigma=-\frac{S_k}{A_k\,t},\
\psi=-\frac{P_k}{A_k\,t},\ V(\Phi)=\frac{B_k^2}{A_k^2\,t^2}\: .
\end{equation}
The stability analysis of the critical
points will give the stability of the above exact solutions.

\subsection{Stability analysis of the singular points}

\subsubsection{The Critical Point $\boldsymbol{P_1}$}

The linearization of ($*$) around the critical point $P_1:
(S,U,P)=(-(2 \sqrt{3})^{-1},$ $0,0)$ is diagonal and has eigenvalues
$\varepsilon_{S,P}=-1/2$ and $\varepsilon_U=1/2$. So $P_1$ is unstable
both to the future and to the past. Starting around $P_1$, the
critical point can never be reached by any solutions of the dynamical
system. 

Using eq. \eqref{BackTr}, the critical point $P_1$ will correspond to
the unstable solution
\begin{equation}
\theta=\frac{2}{t},\ \sigma=-\frac{1}{\sqrt{3}\ t},\ \psi=0,\ V(\Phi)=0\ .
\end{equation}
%

\subsubsection{The Critical Point $\boldsymbol{P_2}$}\label{sssecP2}

In $P_2$ the linearization of ($*$) has the eigenvalues
$\varepsilon_1=-1+\lambda^2/6$ (twice) and
$\varepsilon_2=-2/3+\lambda^2/3$, so the critical point is
asymptotically stable (with respect to all the system ($*$) and for
$\tau \to \infty$) as long as 
$\lambda<\sqrt{2}$ and unstable for $\sqrt{2}<\lambda<\sqrt{6}$.

Since $P_2$ lies on the ellipsoid which separates the KS from the
BIII domain, its neighborhood intersects all three types of
universe. More precisely, starting from a point $P$, near $P_2$, defined by
\begin{equation}
S=\delta S,\ U=\sqrt{\frac{6-\lambda^2}{18}}+\delta U,\
P=\frac{\lambda}{3}+\delta P
\end{equation}
the constraints equation \eqref{transSysConst} expanded to first order
tells us to which universe $P$ belongs since we have
\begin{equation}
\sqrt{\frac{2(6-\lambda^2)}{9}}\,\delta U+\frac{\lambda}{3}\,\delta P =
-\frac{k}{b^2}\ .
\end{equation}
Depending from which point we start the integration, the solution can
be in either type of universe.

As mentioned earlier, the Bianchi~I ($k=0$) metric has to be considered
very carefully because the system is constrained by \eqref{transSysConst}
to the two-dimensional submanifold $W$.
The tangential space to $W$ in the point $P_2$ is given by
\begin{equation}
T_{P_2} W = \{\boldsymbol{X} \in R^3: \boldsymbol{X} \perp \nabla f(P_2)\}\ .
\end{equation}
But $\nabla f(P_2)=(0,\sqrt{2(6-\lambda^2)}/3,\lambda/3)$ is an
eigenvector for the eigenvalue $\varepsilon_2$, and so $T_{P_2}W$
coincides with the eigenspace to the eigenvalue $\varepsilon_1$. That
is, $W$ is the stable submanifold through $P_2$ for $\lambda<\sqrt{6}$
and consequently $P_2$ is in all this range an asymptotically stable
critical point for Bianchi~I solutions, contrary to the Bianchi~III and
Kantowski-Sachs solutions. This particular behavior for the BI case
was overseen in Ref.\cite{Bur:88}. 

For the special values $\lambda=\sqrt{2}$ and $\lambda=\sqrt{6}$ at
least one eigenvalue vanishes and the linearization does not contain
enough information to determine the stability of $P_2$. By projecting
the system ($*$) on a center manifold through $P_2$ \cite{Ama:83,
Arr:90} we find that $P_2$ is unstable for $\lambda=\sqrt{2}$.
In the case $\lambda=\sqrt{6}$ the center
manifold is two dimensional and therefore the stability analysis is
much more difficult. We did not found the asymptotical properties of
the solutions around this singular point for this value of $\lambda$.

The critical point $P_2$ transforms back to the exact solution
\begin{equation}
\theta=\frac{6}{\lambda^2\ t},\ \sigma=0,\ \psi=\frac{2}{\lambda\ t},\
V(\Phi)=\frac{2(6-\lambda^2)}{\lambda^4\ t^2}\ .
\label{solp2}
\end{equation}
 For $\lambda < \sqrt{2}$ this
solution violates the dominant energy condition ($\psi^2 - V <0 $) and
thus inflation occurs. This can be directly seen since from
the above equation we obtain
\begin{equation}
a=a_0 t^{2/\lambda^2} 
\end{equation}
For the asymptotically stable range ($\lambda<\sqrt{2}$ for KS, BIII
and $\lambda<\sqrt{6}$ for BI) of $P_2$
the solution is also asymptotically stable in the sense that solutions
which start near it converge to it.

\subsubsection{The Critical Point $\boldsymbol{P_3}$}

The eigenvalues of the linearization of ($*$) in $P_3$ are
\begin{equation}
\varepsilon_1=-\frac{\lambda^2+2}{2(\lambda^2+1)},\
\varepsilon_{2,3}=\frac{-(\lambda^2+2) \pm \sqrt{(18-7
\lambda^2)(\lambda^2+2)}}{4(\lambda^2+1)}\ .
\end{equation}
For $\lambda<\sqrt{2}$, all eigenvalues are real, $\varepsilon_2$ and
$\varepsilon_3$ have different signs, so $P_3$ is unstable. For
$\lambda=\sqrt{2}$, $P_3$ coincides with $P_2$ so we conclude as before
that the critical point is unstable. In the range $\sqrt{2}<\lambda \leq
\sqrt{18/7}$ all eigenvalues are real and negative, while for
$\lambda>\sqrt{18/7}$ $\varepsilon_2$ and $\varepsilon_3$ become complex
with the same negative real part. So $P_3$ is asymptotically stable for
$\lambda>\sqrt{2}$.

Again we find the following exact solution after the pull-back of
$P_3$
\begin{equation}
\theta=\frac{2(\lambda^2+1)}{\lambda^2\ t},\
\sigma=\frac{2-\lambda^2}{\sqrt{3}\ \lambda^2\ t},\
\psi=\frac{2}{\lambda\ t},\ V(\Phi)=\frac{2(\lambda^2+2)}{\lambda^4\
t^2}\ ,
\end{equation}
which is asymptotically stable for $\lambda>\sqrt{2}$.

\subsubsection{The Critical Points of $\boldsymbol{\Sigma}$}

Since the critical points in $\Sigma$ are on the ellipsoid $E$, 
we find as
before (section \ref{sssecP2}) that there are points belonging to all
three type of universe near $\Sigma$.

For the linearization of ($*$) around a point $P_\Sigma={S_0,0,P_0}$
of $\Sigma$ we find the eigenvalues
\begin{equation}
\varepsilon_1=0,\ \varepsilon_2=1-\frac{\lambda}{2}\,P_0,\
\varepsilon_3=\frac{2}{3}(2+\sqrt{3}\,S_0)\ .
\end{equation}
If $P_0>2/\lambda$, $\varepsilon_2$ and $\varepsilon_3$ have different
signs and the critical point is unstable both to the future and to the
past. 

If $P_0<2/\lambda$, $\varepsilon_2$ and $\varepsilon_3$ are both
positive, so $P_\Sigma$ is unstable to the future. The stability for
$\tau \to -\infty$ is determined by the behavior of the system on a
center manifold \cite{Ama:83, Arr:90}. It is easy to see that $\Sigma$
itself is a center manifold and that the restriction of ($*$) to
$\Sigma$ is just the trivial system
\begin{equation}
S'=0,\ U'=0,\ P'=0\ .
\end{equation}
That means that for $\tau \to -\infty$ $P_\Sigma$ is an attractor. 
It his also easy to see that for $\lambda < \sqrt{6}$  $\Sigma$ as a whole is a
past-attractor for ($*$). Indeed, $\varepsilon_2$ is then positive for
all points $P_\Sigma$.

The pull-back of points out of $\Sigma$ gives an exact
solution of the form
\begin{equation}
\theta=\frac{1}{t},\ \sigma=\frac{\sigma_0}{t},\ \psi=\frac{\psi_0}{t},\
V(\Phi)=0\ ,
\end{equation}
with the condition $\frac{2}{3}-2\,\sigma_0^2-\psi_0^2=0$.

\subsection{Summary}

%
The cases with $\lambda=0$ (constant potential) can be treated
analogously. For the critical points $P_2$ and $P_3$ this has been done
by Burd and Barrow \cite{Bur:88} while for $P_1$ and $\Sigma$ the
results founded are still valid since the asymptotic behavior does not
depend on $\lambda$. 

The results of this section are summarized in table 1 (isolated
critical points) and table 2 (family $\Sigma$). They show for
every critical point the type of the corresponding exact solution
(BI/III: Bianchi~I/III or K-S: Kantowski-Sachs), its Isotropization (Is)
and inflation (In) (e: exponential inflation, p: power-law inflation or
-: no inflation), and the stability (St) of the critical points
($\uparrow$: asymptotic future-stable, $\dagger$: asymptotic
future-stable for Bianchi~I solutions and unstable otherwise,
$\downarrow_\Sigma$: past-convergence to $\Sigma$ and -: unstable).

From table 1, it follows that for $\lambda< \sqrt{2}$ the
unique asymptotical stable solution is represented by the point
$P_2$. This solution describes an inflationary and isotropic universe.

For $\sqrt{2}<\lambda<\sqrt{6}$, isotropy can be reached without
inflation if we restrict ourselves to the BI universe. This fact is
compatible with a Collins and Hawking result \cite{Col:73}, which
states that for 
ordinary matter, within the BI universe, isotropy can be reached without
inflation. Indeed, for $\sqrt{2}<\lambda<\sqrt{6}$ the equation of
state of the matter field corresponding to the $P_2$ solutions is
given by 
\begin{equation}
p=\omega\rho
\end{equation}
with $\omega\in ]-1/3,1[$ and contains the ordinary matter case 
($0\leq\omega<1$).

For $\lambda>\sqrt{6}$, the unique asymptotical stable solution is
represented by the point $P_3$. This BIII solution describes an
anisotropic universe.

There is also a unique attractor in the past: the $\Sigma$
manifold. As a consequence, we must start the numerical integration
near $\Sigma$ for having all the history of the evolution of the model.

The next section will be devoted to the numerical integration.

\section{Numerical Results}\label{secNUM}
In this section we will follow numerically the evolution of the system
\eqref{ed1}-\eqref{ed4}. We have seen in section
\ref{secSTU} that the points of the surface $\Sigma$ are the only
past-attractor of the 
system, so it seems obvious to study the development of solutions
starting near the corresponding exact solution
\begin{equation}
\theta = \frac{1}{t},\ \sigma = \sigma_0 \frac{1}{t},\ \psi = \psi_0
\frac{1}{t},\ V(\Phi) = 0
\end{equation}
with $t\rightarrow 0$ and with $3\sigma_0^2+3/2\psi_0^2=1$.  
Since these exact solutions are in the BI universe, 
the numerical solutions starting near $\Sigma$ can be in any three
type of universe (e.g. BI, BIII or KS).

To distinguish the different future-attractors it is convenient not to plot the
variables $\theta$ and $\sigma$ themselves (they always vanish as $1/t$)
but the quantities $\theta \cdot t$ and $\sigma \cdot t$ which converge
to constant values at the critical points.

\subsection{Kantowski-Sachs Solutions}

For the initial asymptotic behavior 
\begin{equation}\label{AZ}
\theta(t_0)=\frac{1}{t_0},\
\sigma(t_0)=\frac{\sigma_0+\delta\sigma}{t_0},\ u(t_0)=\frac{\delta
u}{t_0},\ \psi(t_0)=\frac{\psi_0+\delta\psi}{t_0}\ ,
\end{equation}
we will have a Kantowski-Sachs solutions if the following inequality
is satisfied
\begin{equation}
2 \sigma_0 \delta\sigma+\psi_0 \delta\psi>0\ .
\end{equation}
Fig.~2 shows the results for some values of $\lambda$ less then
the critical value $\sqrt{2}$. The solution clearly converges towards
the exact solution corresponding to the critical point $P_2$. The
corresponding phase portraits are given in fig.~3.

The development of a solution with $\lambda>\sqrt{2}$ is shown in
fig~4. The solution approaches again the (now unstable)
attractor $P_2$, is repelled and finally converges towards the exact
solution corresponding to the attractor $P_3$.

\subsection{Bianchi~III Solutions}

For a Bianchi~III solution of the form \eqref{AZ} the integration
constants have to satisfy the condition
\begin{equation}
2 \sigma_0 \delta\sigma+\psi_0 \delta\psi<0\ .
\end{equation}
The solutions with $\lambda<\sqrt{2}$ (fig.~5) initially
tend to the Bianchi~III solution corresponding to $P_1$. But since this
is an unstable attractor the system finally evolves towards the solution
$P_2$. Fig.~6 shows the corresponding phase portraits for
these cases. 

For $\lambda>\sqrt{2}$ the system again approaches the unstable
$P_1$-solution in the beginning, is repelled and converges towards the
$P_3$-solution which is the only attractor in this range of
$\lambda$-values (fig.~7, phase portraits:
fig.~8).

\subsection{Bianchi~I Solutions}

For a Bianchi~I solution, the perturbations of the initial state
\eqref{AZ} have to satisfy the equation
\begin{equation}
2 \sigma_0 \delta\sigma+\psi_0 \delta\psi=0\ .
\end{equation}

For Bianchi~I solutions, the exact solution corresponding to $P_2$ is
attractive not only for $\lambda<\sqrt{2}$ but for the whole range for
which $P_2$ is defined, that is for $\lambda<\sqrt{6}$
(fig.~9, phase portraits: fig.~10).
The numerical calculation confirms our analysis
for $\sqrt{2}<\lambda<\sqrt{6}$: the solutions still
isotropize but are no longer inflationary as one can see from
fig.~11 where we have plotted $\psi^2-V(\Phi)$ which
violates the dominant energy condition when negative.

For $\lambda>\sqrt{6}$ we have again the solution to $P_3$ as the only
attractor. An example for this case is shown in fig.~12. The
system quickly evolves towards the now unstable solution $P_2$, is
repelled and converges towards the solution to $P_3$.

\section{Conclusions}\label{secCON}
We have studied the Einstein-Klein-Gordon (EKG) equations for a convex
positive potential in a Bianchi I, Bianchi III and a Kantowsky-Sachs
universe. 

After analyzing the inherent properties of the equations, it was shown
in section \ref{secINH} why a detailed analysis of the 
solutions of the EKG equations was needed for a vanishing potential at
infinity. By taking an exponential potential ($V=V_0
e^{-\lambda\varphi}$) it was shown for which values of $\lambda$
inflation and/or isotropy where reached asymptotically. We recovered
the results of Ref. \cite{Bur:88} but also found new asymptotical
behaviors and new exact solutions represented by the singular points
$P_1$ and the submanifold $\Sigma$. 

We also found that for some values of $\lambda$ isotropy can be
reached without inflation in a Bianchi~I universe. But when inflation
occurs ($\lambda<\sqrt{2}$) then isotropy is always reached. 
We also integrated
the equations numerically to obtain the all history of the evolution
of the model. All the founded 
asymptotical behavior were confirmed numerically.   

\section*{Aknowledgments}
We thank N. Deruelle for very useful discussions and for carefully
reading the manuscript. We also thank N. Straumann to attract our
attention to this problem.

\pagebreak
\section*{Caption}

\noindent Table 1: 

\noindent Summary of results for isolated critical points

\vskip 0.5cm
\noindent Table 2: 

\noindent Summary of results for critical points in $\Sigma$

\vskip 0.5cm 
\noindent Figure 1: 

\noindent Region of the space with coordinates ($\theta$, $\sigma$,
$z$, $\psi$) where the solutions lie as 
function of $k$. The solutions for the B-I case are restricted in a
submanifold of codimension 1, the ``lightcone''.

\vskip 0.5cm
\noindent Figure 2: 

\noindent Evolution of Kantowski-Sachs solutions with $\lambda < 
\protect\sqrt{2}$.

\vskip 0.5cm
\noindent Figure 3: 

\noindent The corresponding phase portraits for the KS case with $\lambda < 
\protect\sqrt{2}$.

\vskip 0.5cm
\noindent Figure 4: 

\noindent Evolution of a Kantowski-Sachs solution with 
$\lambda> \protect\sqrt{2}$.

\vskip 0.5cm
\noindent Figure 5: 

\noindent Evolution of Bianchi~III solutions with 
$\lambda<   \protect\sqrt{2}$.

\vskip 0.5cm
\noindent Figure 6: 

\noindent The corresponding phase portraits for the BIII case with $\lambda < 
\protect\sqrt{2}$.

\vskip 0.5cm
\noindent Figure 7: 

\noindent Evolution of Bianchi~III solutions with 
$\lambda> \protect\sqrt{2}$.

\vskip 0.5cm
\noindent Figure 8: 

\noindent The corresponding phase portraits for the BIII case with $\lambda > 
\protect\sqrt{2}$.

\vskip 0.5cm
\noindent Figure 9: 

\noindent Evolution of Bianchi~I solutions with 
$\lambda< \protect\sqrt{6}$.

\vskip 0.5cm
\noindent Figure 10: 

\noindent The corresponding phase portraits for the BI case with $\lambda < 
\protect\sqrt{6}$.

\vskip 0.5cm
\noindent Figure 11: 

\noindent Inflation can only occur when the dominant energy condition
is violated that is when $(\psi^2-V) t^2$ is negative.  
$(\psi^2-V) t^2$ is plotted for the corresponding numerical solutions
described by fig. 9 and fig. 10.

\vskip 0.5cm
\noindent Figure 12: 

\noindent Evolution of a Bianchi~I solution with 
$\lambda> \protect\sqrt{6}$.
\pagebreak
\begin{table}[htb] 
\begin{center}
\begin{tabular}{|c||c|c|c|c||c|c|c|c||c|c|c|c|}
\hline
& \multicolumn{4}{c||}{$P_1\ $} & \multicolumn{4}{c||}{$P_2\ $} &
\multicolumn{4}{c|}{$P_3\ $} \\  
\hline
$\lambda$ & \scs Type & \scs Is & \scs In & \scs St & \scs Type & \scs Is
& \scs In & \scs St & \scs Type & \scs Is & \scs In & \scs St \\ 
\hline
\hline
\scs $\lambda=0$ &  B III &  $-$ &  $-$ &  $-$ &  B I &  $+$ &  e &
$\uparrow$ &  K-S &  $-$ &  e &  $-$ \\ 
\hline
\scs $0<\lambda<\sqrt{2}$ &  B III &  $-$ &  $-$ &  $-$ &  B I &  $+$ &
p &  $\uparrow$ &  K-S &  $-$ &  p &  $-$ \\ 
\hline
\scs $\lambda=\sqrt{2}$ &  B III &  $-$ &  $-$ &  $-$ &  B I &  $+$ &
$-$ &  $\dagger$ &  B I &  $+$ &  $-$ &  $\dagger$ \\ 
\hline
\scs $\sqrt{2}<\lambda<\sqrt{6}$ &  B III &  $-$ &  $-$ &  $-$ &  B I &
$+$ &  $-$ &  $\dagger$ &  B III &  $-$ &  $-$ &  $\uparrow$ \\ 
\hline
\scs $\lambda=\sqrt{6}$ &  B III &  $-$ &  $-$ &  $-$ &  B I &  $+$ &
$-$ &  ? &  B III &  $-$ &  $-$ &  $\uparrow$ \\ 
\hline
\scs $\lambda>\sqrt{6}$ &  B III &  $-$ &  $-$ &  $-$ & 
\multicolumn{4}{c||}{not defined} &  B III &  $-$ &  $-$ &  $\uparrow$ \\
\hline
\end{tabular}
\end{center}
\vskip -\baselineskip
\vskip 0.5cm
\hbox{\hskip 5cm Tab.1}
\vskip -0.5cm
\end{table}
\begin{table}[hbt] 
\begin{center}
\begin{tabular}{|c|c|c||c|c|c|c|}
\hline
& \multicolumn{6}{|c|}{$P: (S_0,0,P_0) \in \Sigma\ $} \\
\hline
$\lambda$ & $P_0$ & $S_0$ &\scs Type & \scs Is & \scs In & \scs St  \\
\hline
\hline
& \scs $P_0=\pm \sqrt{2/3}$ & \scs $S_0=0$ & B I & $+$ & $-$ &
$\downarrow_\Sigma$ \\ 
\cline{2-7}
\raisebox{1.5ex}[0pt]{\scs $\lambda<\sqrt{6}$} & \scs $-\sqrt{2/3}<P_0<\sqrt{2/3}$ & \scs $S_0 \neq 0$ & B I & $-$ & $-$ & $\downarrow_\Sigma$ \\
\hline
& \scs $P_0 = -\sqrt{2/3}$ & \scs $S_0=0$ & B I & $+$ & $-$ &
$\downarrow_\Sigma$ \\ 
\cline{2-7}
\scs $\lambda=\sqrt{6}$ & \scs $-\sqrt{2/3}<P_0<\sqrt{2/3}$ & \scs $S_0
\neq 0$ & B I & $-$ & $-$ & $\downarrow_\Sigma$ \\ 
\cline{2-7}
& \scs $P_0=\sqrt{2/3}$ & \scs $S_0=0$ & B I & $+$ & $-$ & ? \\
\hline
& \scs $P_0=-\sqrt{2/3}$ & \scs $S_0=0$ & B I & $+$ & $-$ &
$\downarrow_\Sigma$ \\ 
\cline{2-7}
& \scs $-\sqrt{2/3}<P_0<2/\lambda$ & \scs $S_0 \neq 0$ & B I & $-$ & $-$
& $\downarrow_\Sigma$ \\ 
\cline{2-7}
\scs $\lambda>\sqrt{6}$ & \scs $P_0=2/\lambda$ & \scs $S_0 \neq 0$ & B I
& $-$ & $-$ & ? \\ 
\cline{2-7}
& \scs $2/\lambda<P_0<\sqrt{2/3}$ & \scs $S_0 \neq 0$ & B I & $-$ & $-$
& $-$ \\ 
\cline{2-7}
& \scs $P_0=\sqrt{2/3}$ & \scs $S_0=0$ & B I & $+$ & $-$ & $-$ \\
\hline
\end{tabular}
\end{center}
\vskip -\baselineskip
\vskip 0.5cm
\hbox{\hskip 5cm Tab.2}
\vskip -0.5cm
\end{table}
\pagebreak
\begin{figure}[!ht]
\begin{center}
\epsfig{file=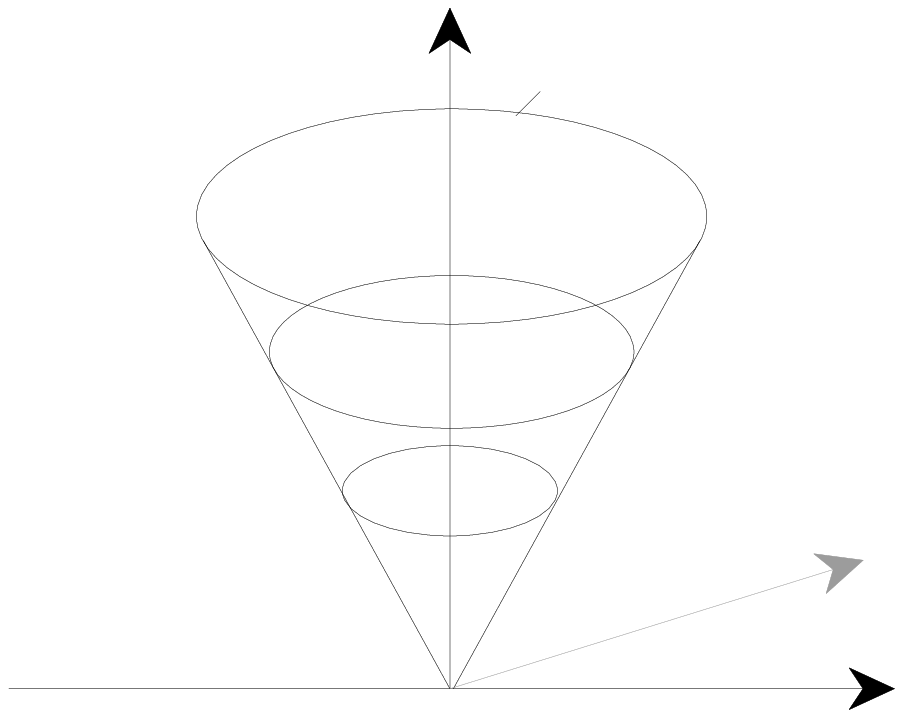,width=12cm}
\end{center}
\vskip -\baselineskip
\vskip -0.7cm
\hbox{\hskip 11cm $\sigma$}
\vskip 0.7cm
\vskip -\baselineskip
\vskip -2cm
\hbox{\hskip 11cm $z,\psi$}
\vskip 2cm
\vskip -\baselineskip
\vskip -8.4cm
\hbox{\hskip 6.5cm $\theta$}
\vskip 8.4cm
\vskip -\baselineskip
\vskip -8.4cm
\hbox{\hskip 8cm $k=0$}
\vskip 8.4cm
\vskip -\baselineskip
\vskip -7cm
\hbox{\hskip 5.3cm $k= -1$}
\vskip 7cm
\vskip -\baselineskip
\vskip -5.5cm
\hbox{\hskip 10.5cm $k=1$}
\vskip 5.5cm
%
%
\vskip -\baselineskip
\vskip 0.5cm
\hbox{\hskip 7cm Fig.1}
\vskip -0.5cm
\end{figure}
%
\begin{figure}[hbt]
\picth{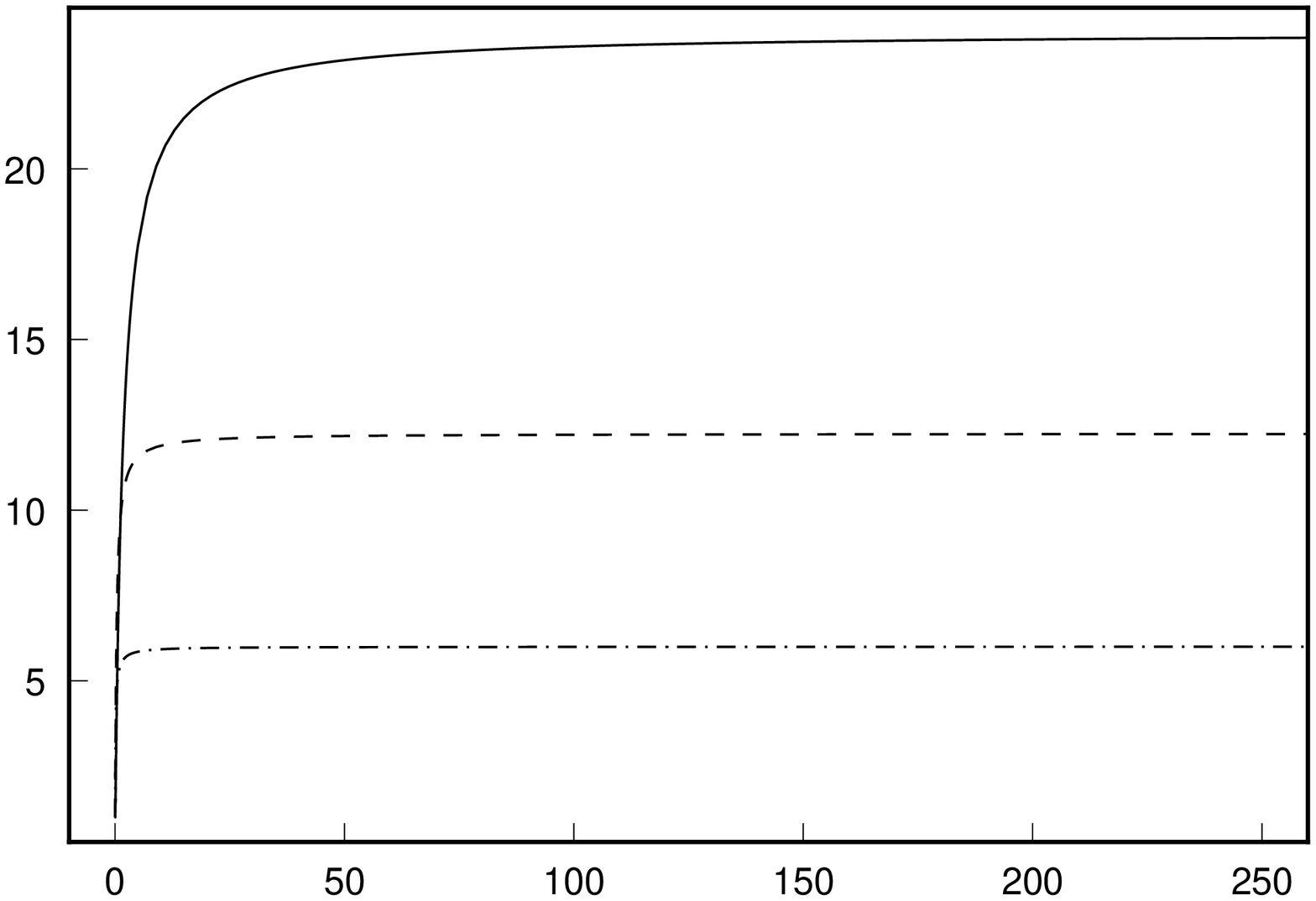}{(2,2.7)}{\lambda=0.5,\sigma_0=-0.5,\psi_0>0}{(2.2,1.5)}
{\lambda=0.7,\sigma_0=-0.2,\psi_0<0}{(2.4,0.8)}
{\lambda=1.0,\sigma_0=0.3,\psi_0<0}{(0,0)}{}
\hfill
\spic{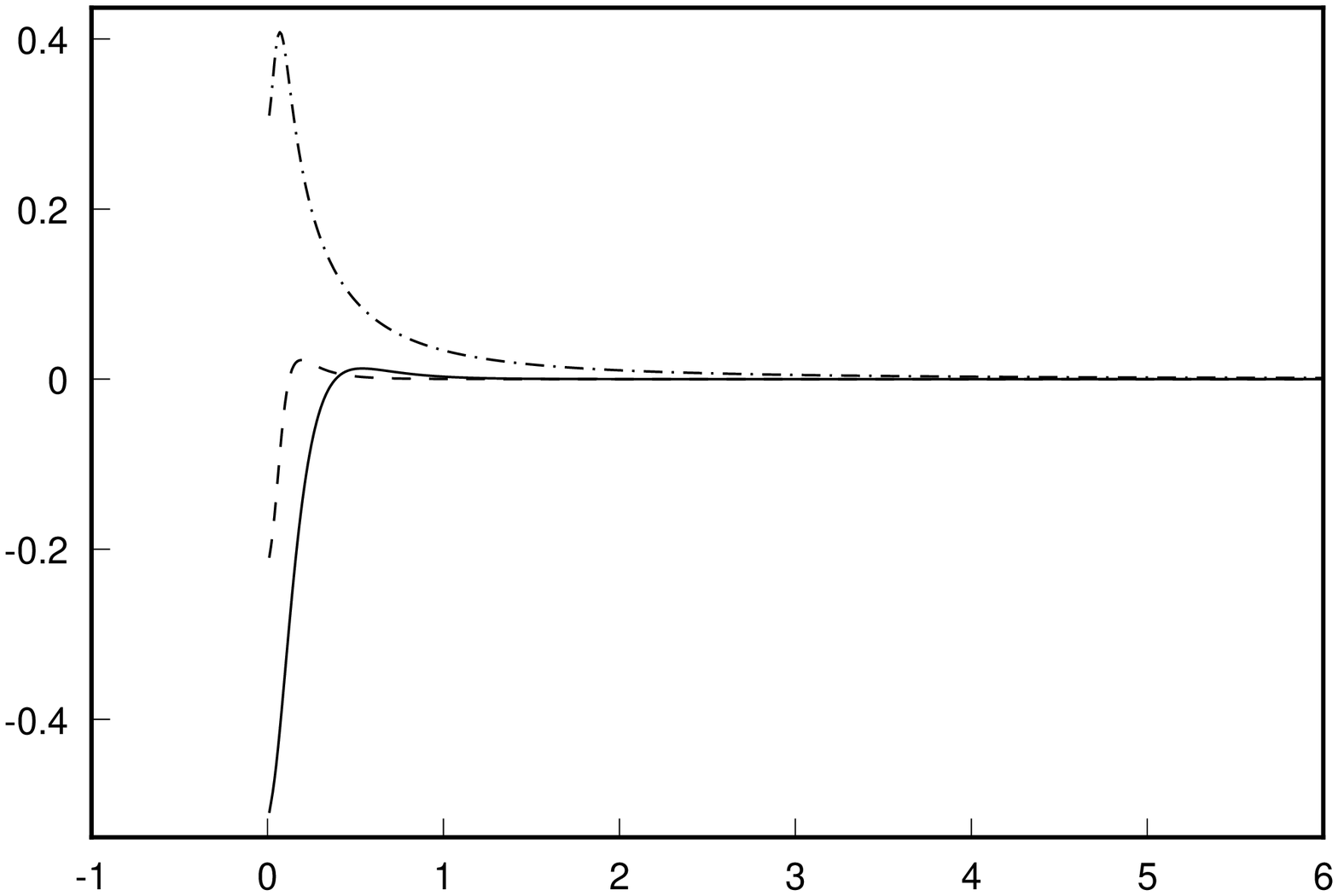}{\sigma \cdot t}\hbox{}
\vskip -\baselineskip
\vskip 0.5cm
\hbox{\hskip 7cm Fig.2}
\vskip -0.5cm
\end{figure}
%
\begin{figure}[hbt]
\begin{center}
\pic{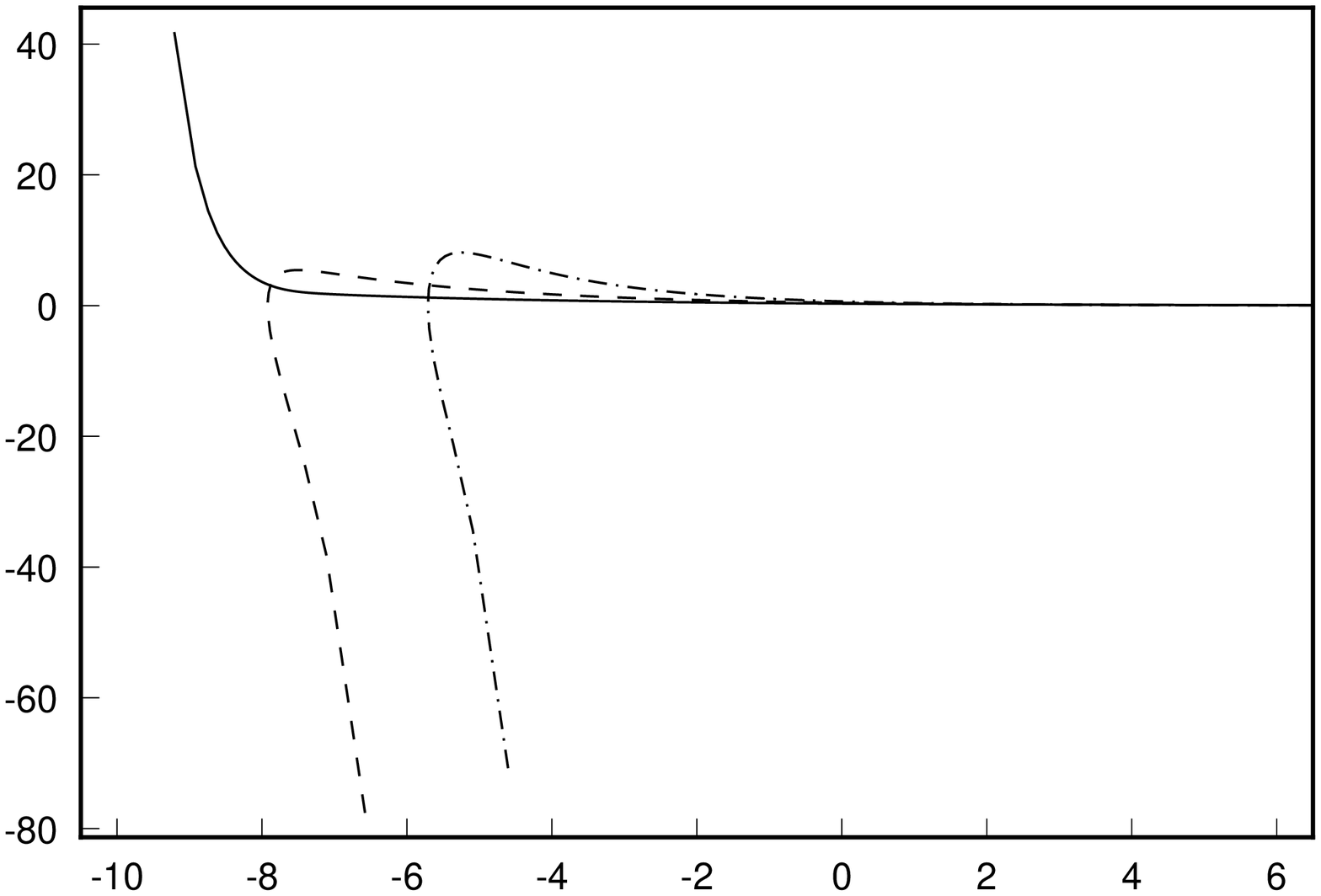}{$\scriptstyle \psi$}{$\scriptstyle \Phi$}
\end{center}
\vskip -\baselineskip
\vskip 0.5cm
\hbox{\hskip 7cm Fig.3}
\vskip -0.5cm
\end{figure}
%
\begin{figure}[hbt]
\picth{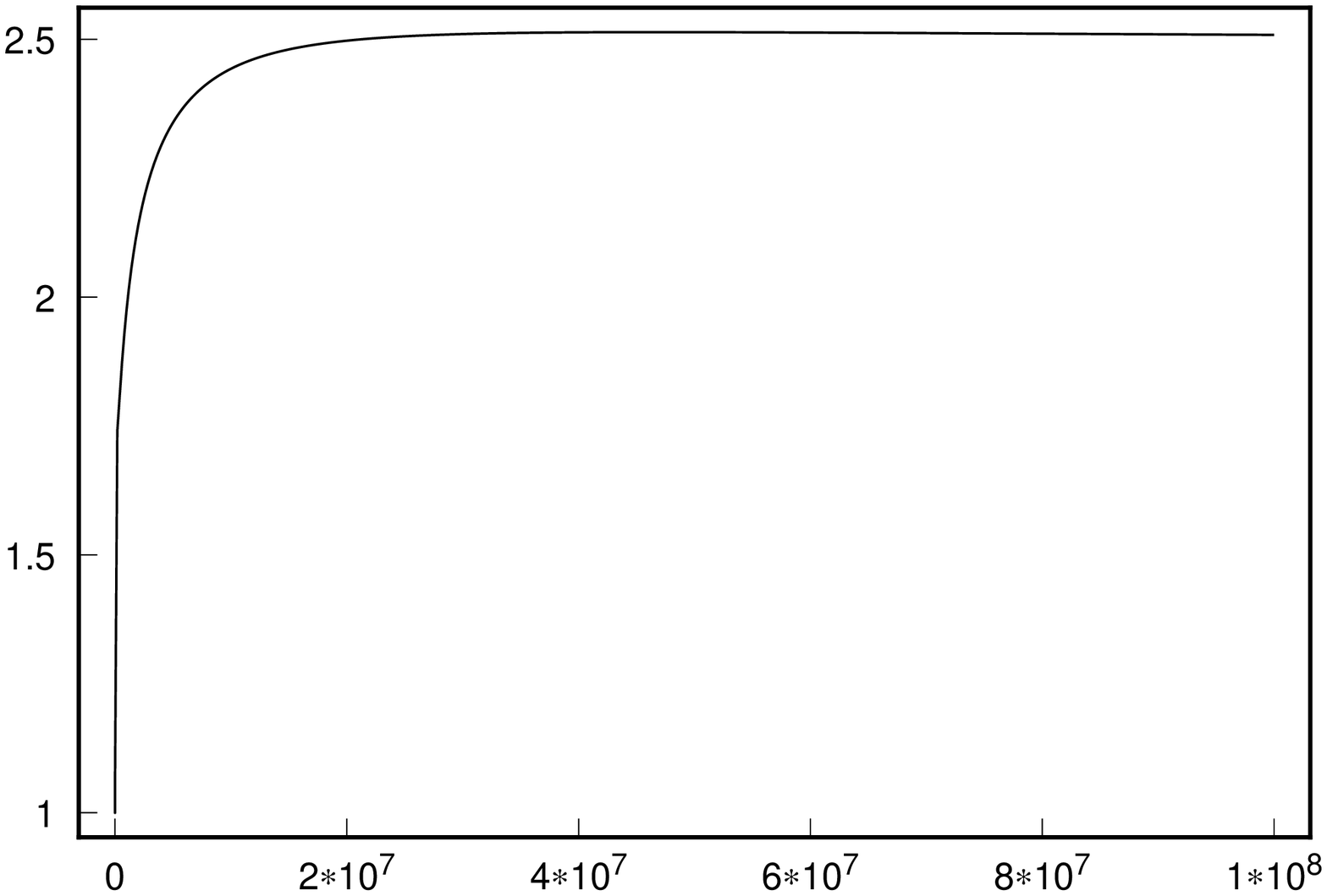}{(2,2.7)}{\lambda=2,\sigma_0=-0.3,\psi_0<0}{(0,0)}{}{(0,0)}{}{(0,0)}{}
\hfill
\spic{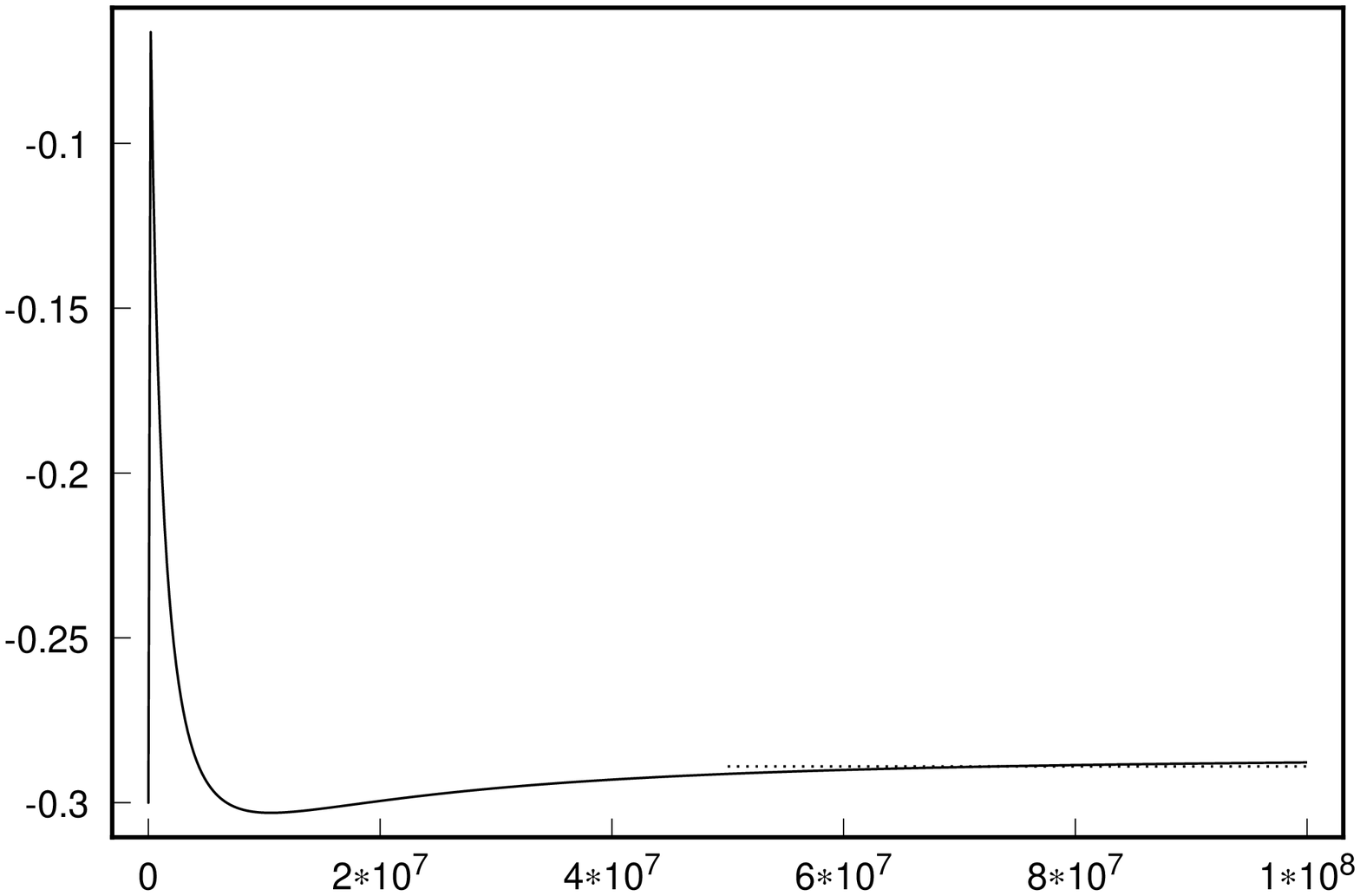}{\sigma \cdot t}\hbox{}
\vskip -\baselineskip
\vskip 0.5cm
\hbox{\hskip 7cm Fig.4}
\vskip -0.5cm
\end{figure}
%
\begin{figure}[ht]
\picth{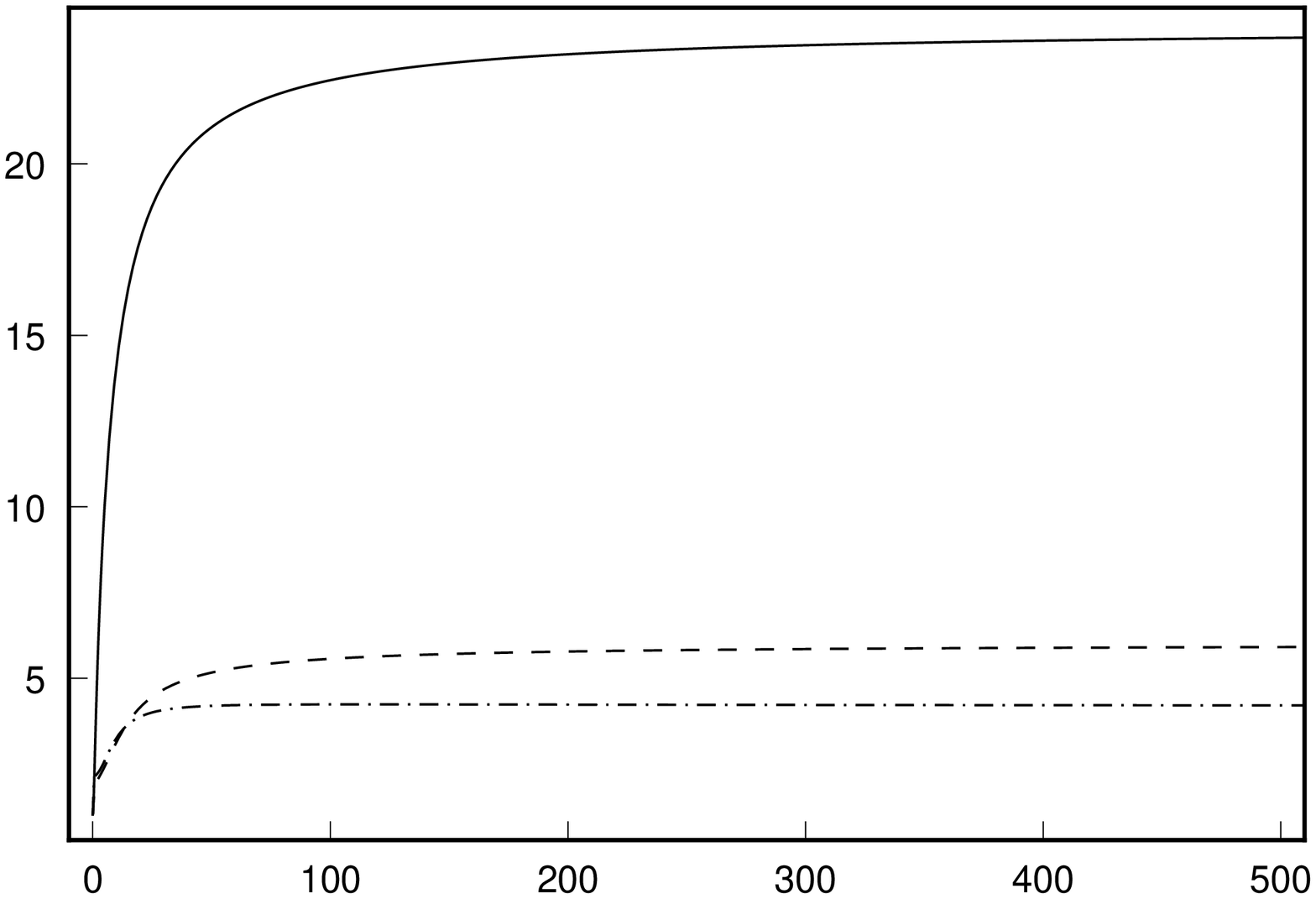}{(2,2.7)}{\lambda=0.5,\sigma_0=-0.4,\psi_0<0}{(2.2,1.3)}%
{\lambda=1,\sigma_0=0,\psi_0>0}{(2.4,0.75)}%
{\lambda=1.2,\sigma_0=0.3,\psi_0>0}{(0,0)}{}
\hfill
\spic{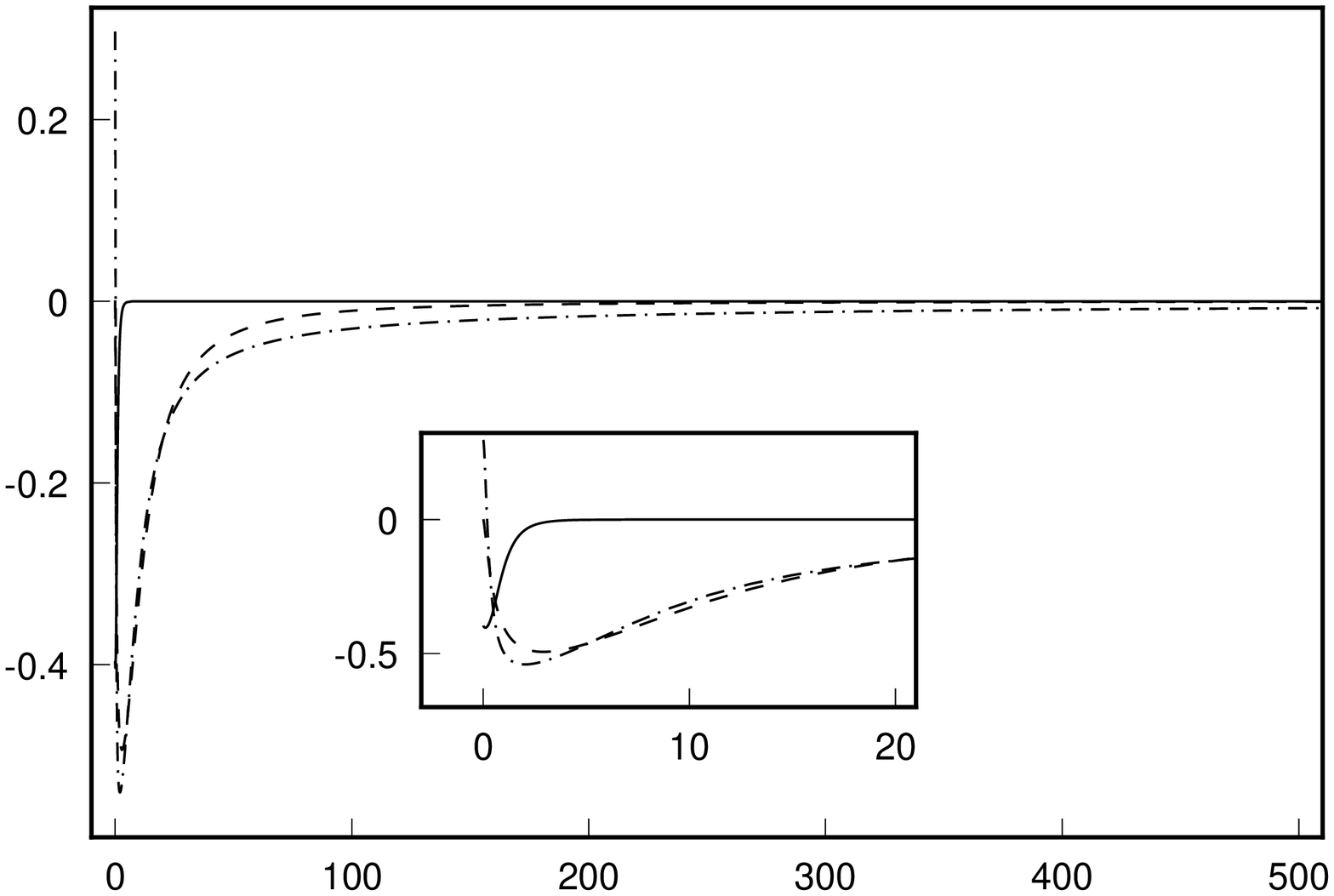}{\sigma \cdot t}\hbox{}
\vskip -\baselineskip
\vskip 0.5cm
\hbox{\hskip 7cm Fig.5}
\vskip -0.5cm
\end{figure}
%
\begin{figure}[ht]
\begin{center}
\pic{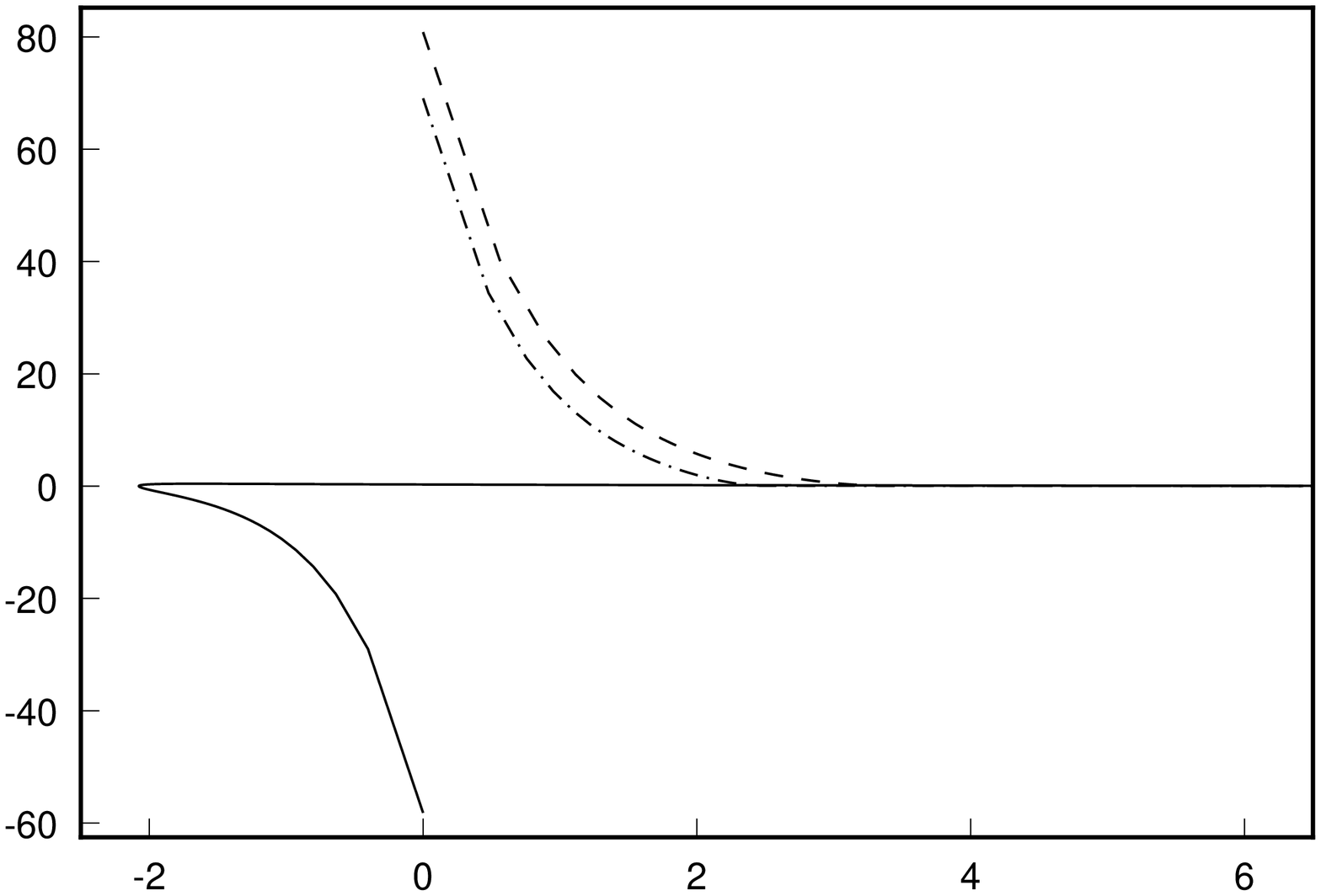}{$\scriptstyle \psi$}{$\scriptstyle \Phi$}
\end{center}
\vskip -\baselineskip
\vskip 0.5cm
\hbox{\hskip 7cm Fig.6}
\vskip -0.5cm
\end{figure}
%
\begin{figure}[ht]
\picth{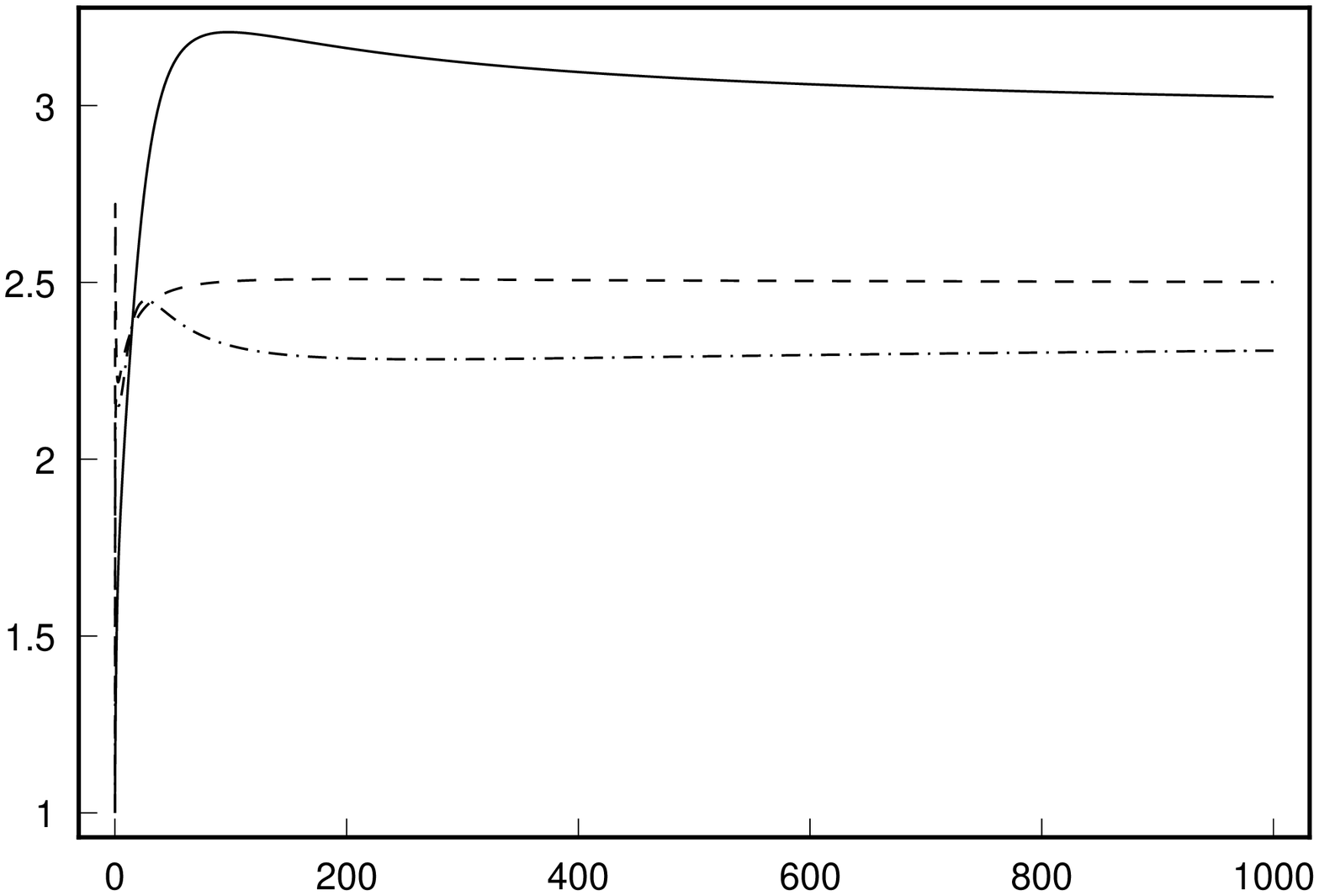}{(2,2.75)}{\lambda=1.5,\sigma_0=-0.3,\psi_0>0}%
{(2.2,2.35)}{\lambda=2,\sigma_0=0.1,\psi_0<0}{(2.4,1.85)}%
{\lambda=2.5,\sigma_0=0.4,\psi_0>0}{(0,0)}{}
\hfill
\spic{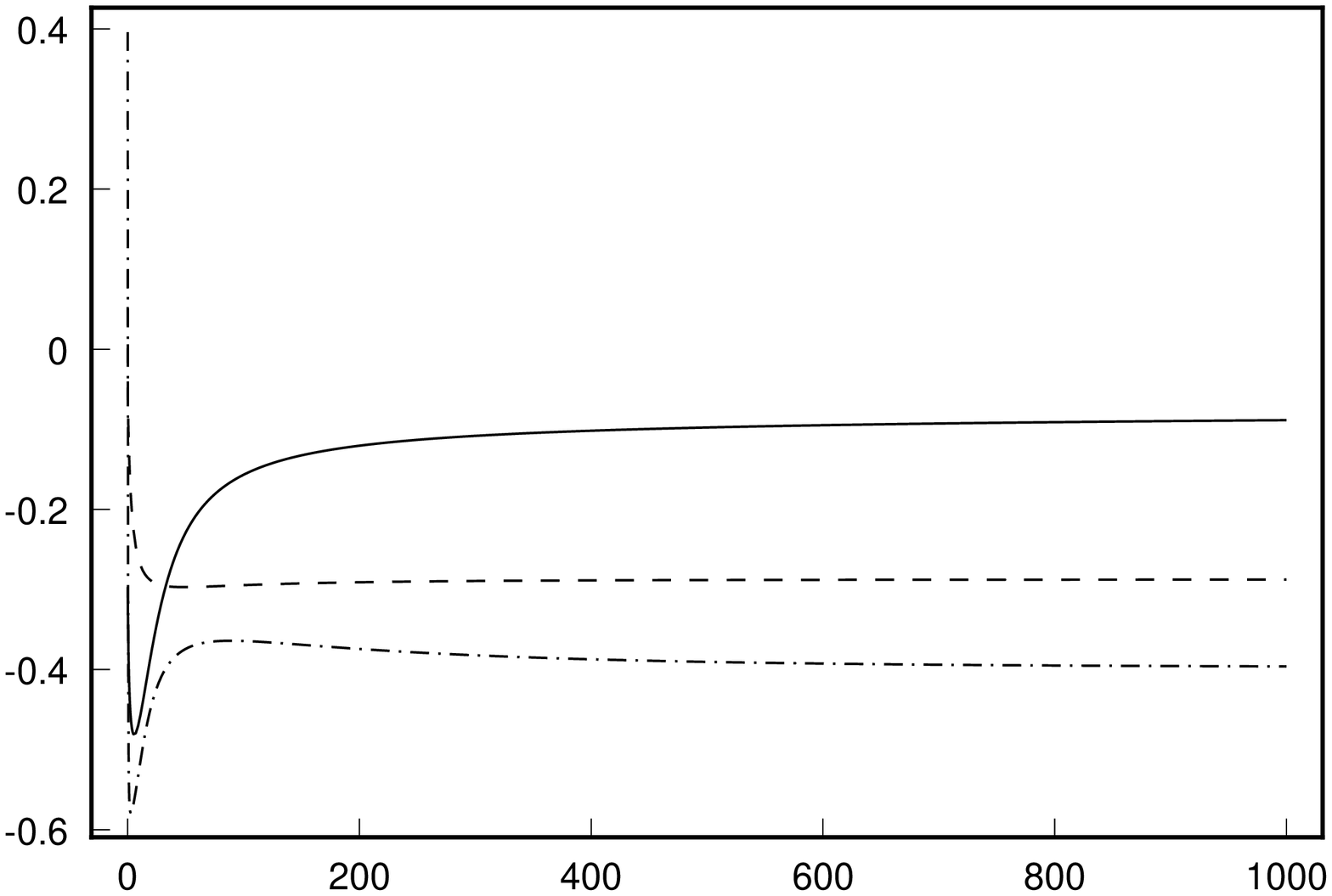}{\sigma \cdot t}\hbox{}
\vskip -\baselineskip
\vskip 0.5cm
\hbox{\hskip 7cm Fig.7}
\vskip -0.5cm
\end{figure}
%
\begin{figure}[hbt]
\begin{center}
\pic{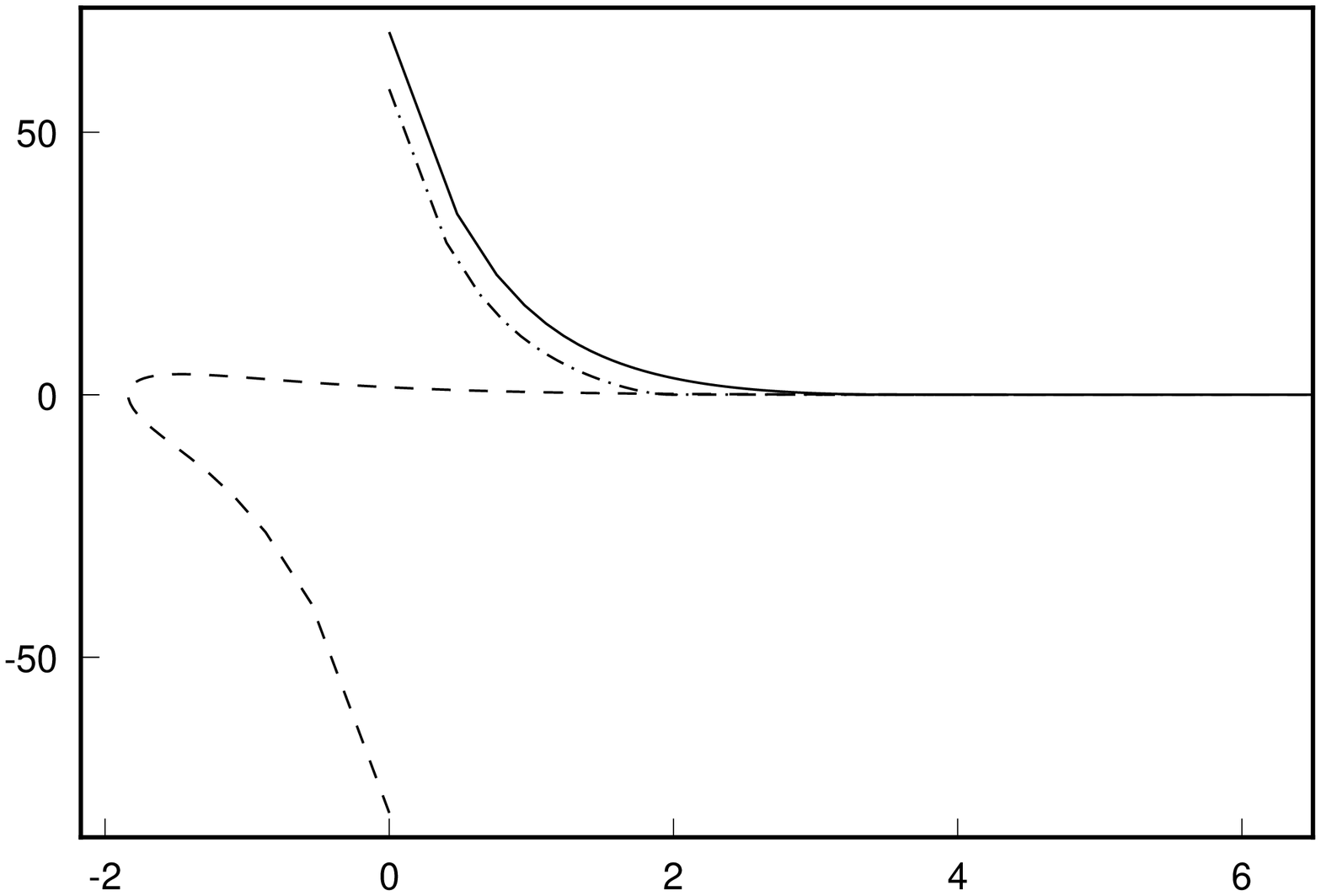}{$\scriptstyle \psi$}{$\scriptstyle \Phi$}
\end{center}
\vskip -\baselineskip
\vskip 0.5cm
\hbox{\hskip 7cm Fig.8}
\vskip -0.5cm
\end{figure}
%
\begin{figure}[hbt]
\picth{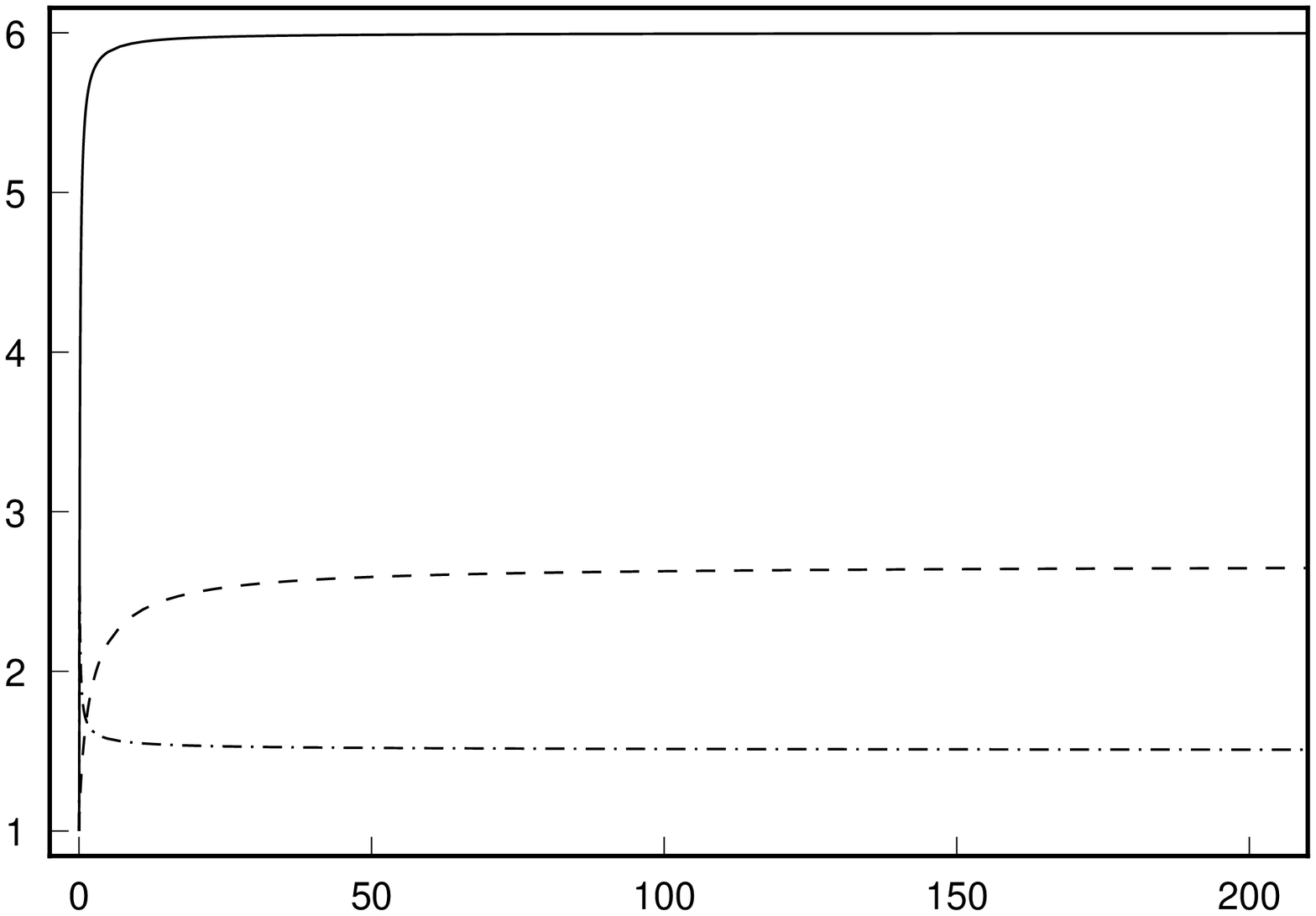}{(2,2.9)}{\lambda=1,\sigma_0=-0.5,\psi_0<0}%
{(2.2,1.2)}{\lambda=1.5,\sigma_0=-0.25,\psi_0>0}{(2.4,0.65)}%
{\lambda=2,\sigma_0=0.3,\psi_0<0}{(0,0)}{}
\hfill
\spic{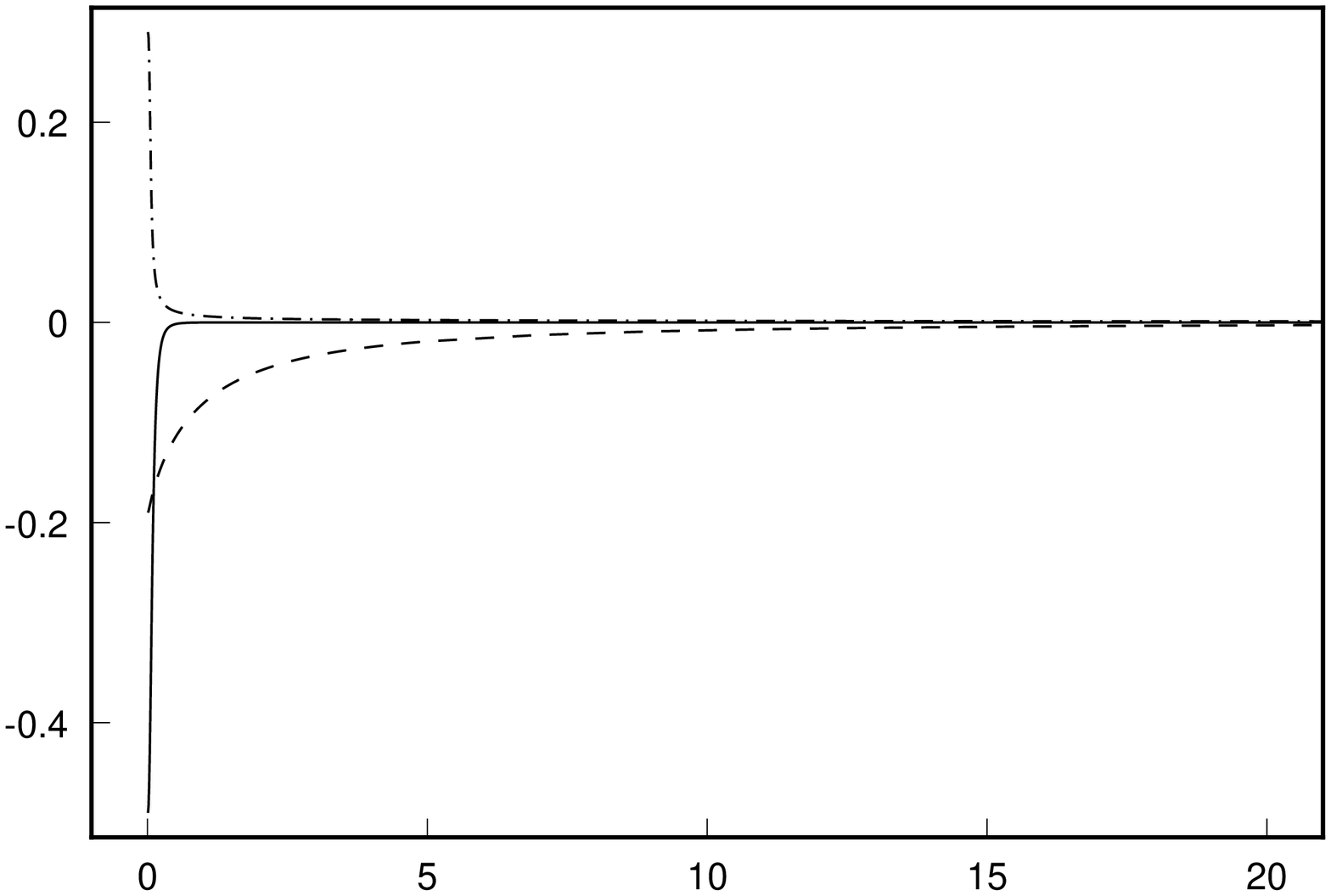}{\sigma \cdot t}\hbox{}
\vskip -\baselineskip
\vskip 0.5cm
\hbox{\hskip 7cm Fig.9}
\vskip -0.5cm
\end{figure}
%
\begin{figure}[hbt]
\begin{center}
\pic{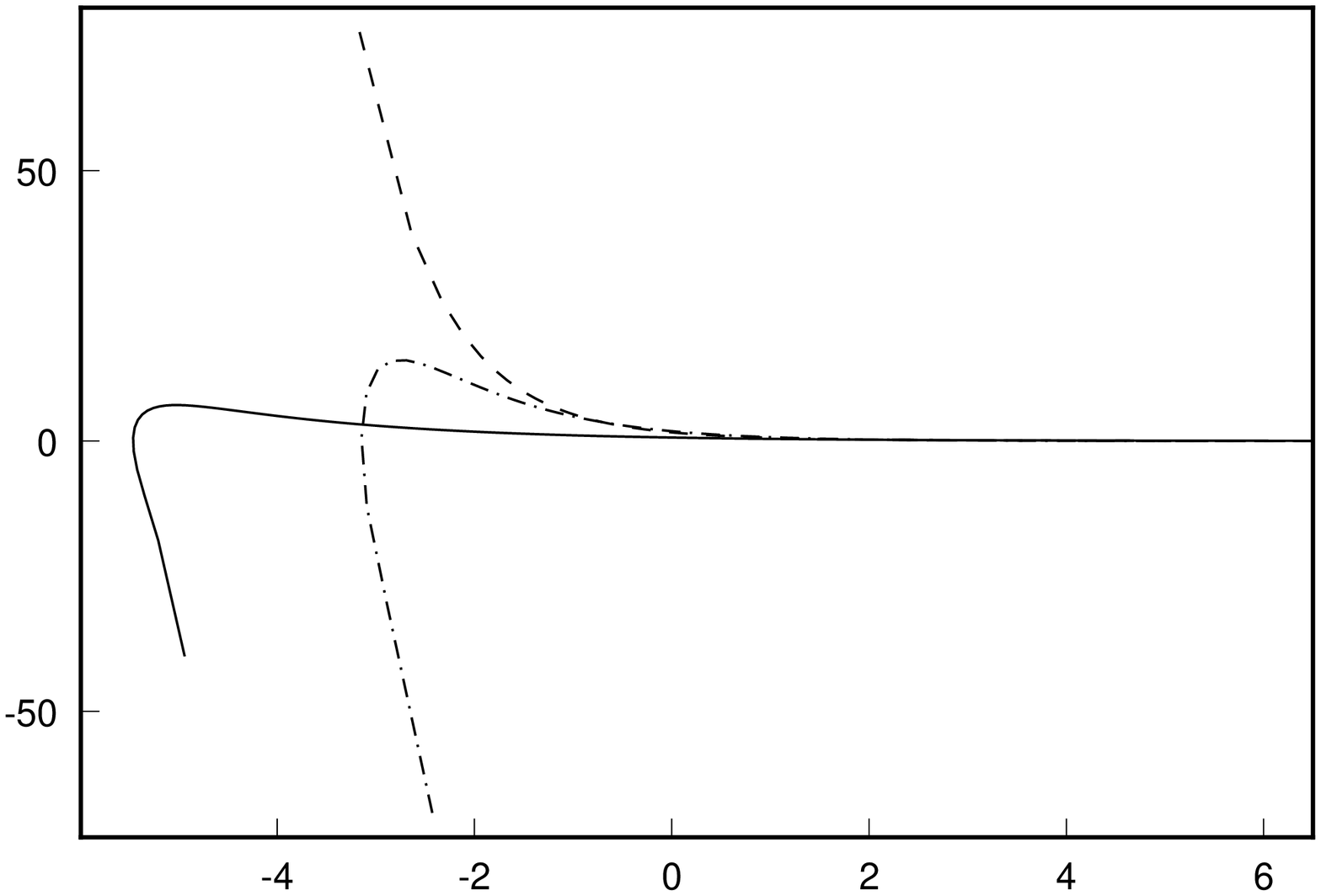}{$\scriptstyle \psi$}{$\scriptstyle \Phi$}
\end{center}
\vskip -\baselineskip
\vskip 0.5cm
\hbox{\hskip 7cm Fig.10}
\vskip -0.5cm
\end{figure}
%
\begin{figure}[hbt]
\begin{center}
\setlength{\unitlength}{1.28cm}
\begin{picture}(5,3.3)
\epsfxsize=5.8cm
\put(0.4,0.1) {\epsfbox{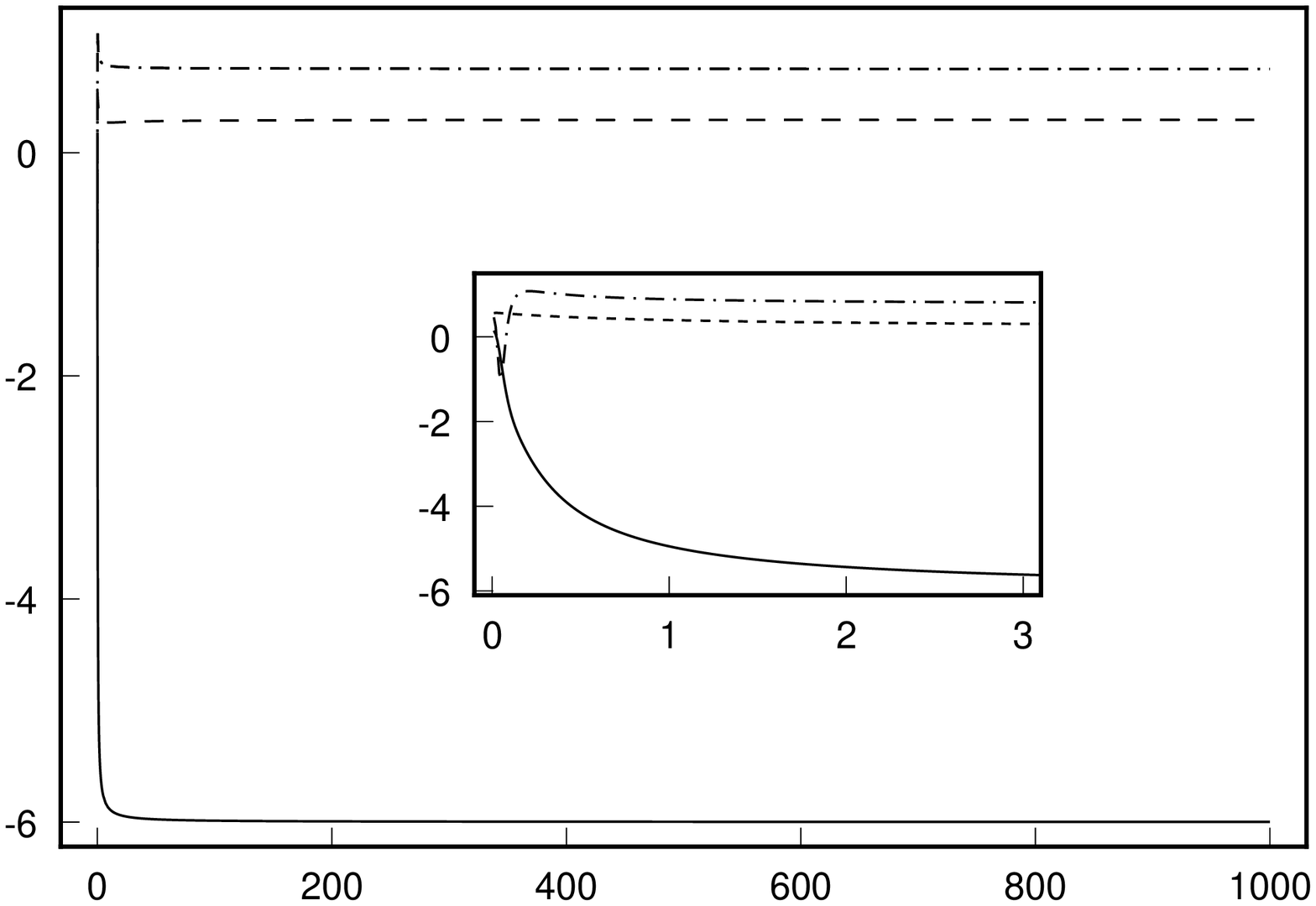}}
\put (0,1.7) {$\scriptscriptstyle (\psi^2-V) t^2$}
\put (2.8,0.1) {$\scriptstyle t$}
\end{picture}
\end{center}
\vskip -\baselineskip
\vskip 0.5cm
\hbox{\hskip 7cm Fig.11}
\vskip -0.5cm
\end{figure}
%
\begin{figure}[hbt]
\picth{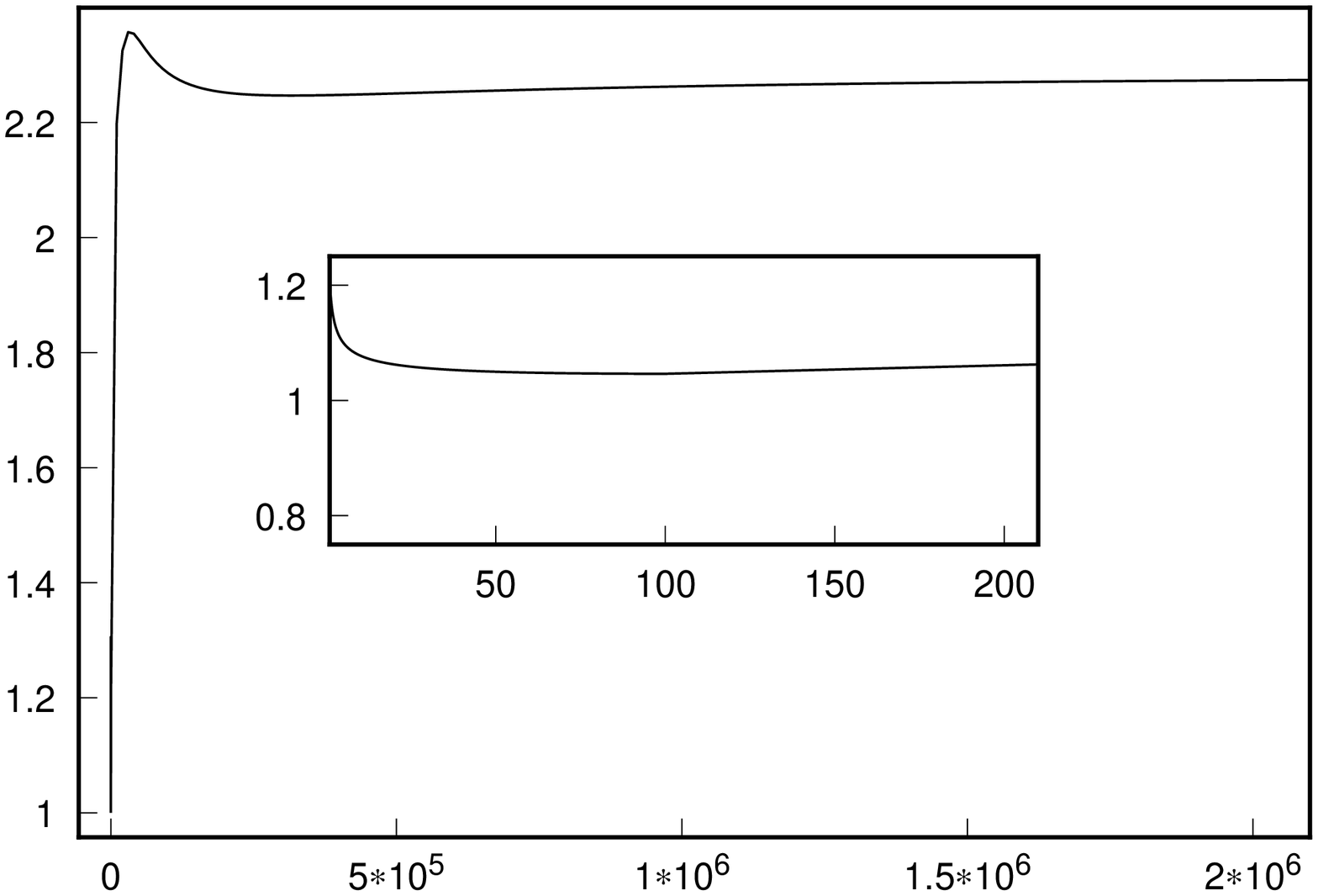}{(2,2.7)}{\lambda=2.7,\sigma_0=0.2,\psi_0<0}%
{(0,0)}{}{(0,0)}{}{(0,0)}{}
\hfill    
\spic{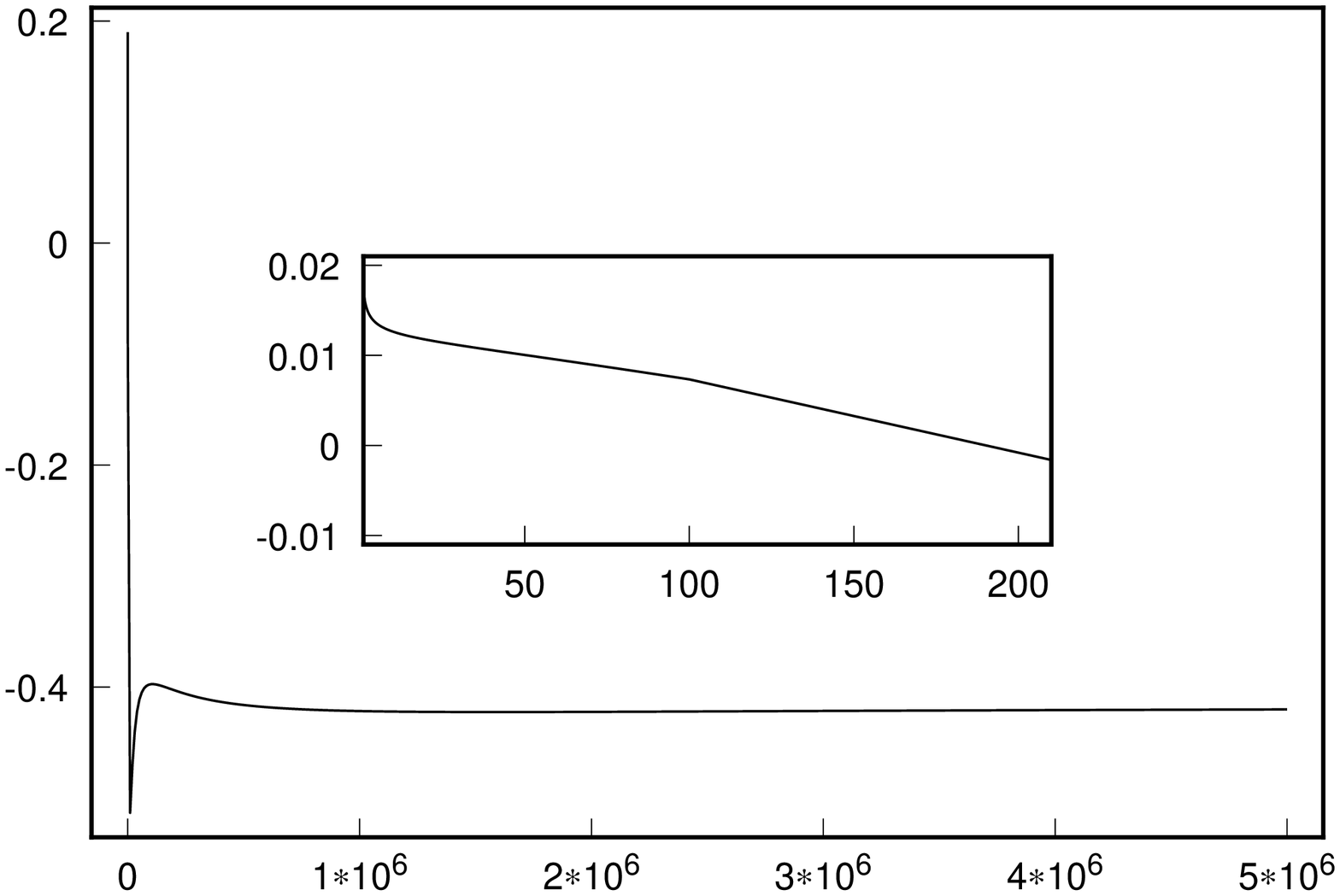}{\sigma \cdot t}\hbox{}
\vskip -\baselineskip
\vskip 0.5cm
\hbox{\hskip 7cm Fig.12}
\vskip -0.5cm
\end{figure}
\end{document}